\documentclass[pra,twocolumn,superscriptaddress]{revtex4-2}
\usepackage{booktabs} 
\usepackage{multirow} 
\usepackage{geometry}
\usepackage{lineno}
\usepackage{siunitx}

\usepackage{graphicx}
\usepackage[dvipsnames]{xcolor}
\usepackage{bm}
\usepackage{physics}
\usepackage{amsthm}
\usepackage{mathtools}
\usepackage[hidelinks]{hyperref}
\usepackage{bbm}
\usepackage{amsfonts}
\usepackage{float}
\usepackage{appendix}
\usepackage[capitalise]{cleveref}
\usepackage[normalem]{ulem}
\theoremstyle{definition}

\newtheorem{rst}{Result}

\begin{document}

\preprint{APS/123-QED}

\title{Experimental Tabletop Petz recovery of a photonic qubit}

\author{Hui Li}
\email{Co-first author (experiment)}
\affiliation{National Laboratory of Solid State Microstructures, College of Engineering and Applied Science, Nanjing University, Nanjing 210093, China}

\author{Jinyan Chen}
\email{Co-first author (theory)}
\affiliation{Centre for Quantum Technologies, National University of Singapore, 3 Science Drive 2, Singapore 117543}

\author{Yue Pan}
\affiliation{National Laboratory of Solid State Microstructures, College of Engineering and Applied Science, Nanjing University, Nanjing 210093, China}

\author{Liang Xu}
\affiliation{National Laboratory of Solid State Microstructures, College of Engineering and Applied Science, Nanjing University, Nanjing 210093, China}

\author{Minjeong Song}
\affiliation{Centre for Quantum Technologies, National University of Singapore, 3 Science Drive 2, Singapore 117543}

\author{Valerio Scarani}
\email{physv@nus.edu.sg}
\affiliation{Centre for Quantum Technologies, National University of Singapore, 3 Science Drive 2, Singapore 117543}
\affiliation{Department of Physics, National University of Singapore, 2 Science Drive 3, Singapore 117542}

\author{Lijian Zhang}
\email{lijian.zhang@nju.edu.cn}
\affiliation{National Laboratory of Solid State Microstructures, College of Engineering and Applied Science, Nanjing University, Nanjing 210093, China}

\date{\today}

\begin{abstract}
    The quantum information lost in open evolutions cannot be fully recovered, but partial recovery is possible. The Petz recovery map guarantees almost optimal recovery, notably if the chosen reference state is close to the real one. This map has been widely used in theoretical studies, but has been the object of only a handful of experimental realisations, typically under a single fixed noise model. In this work, we describe and implement the Petz recovery map for a versatile class of qubit channels with tunable decoherence and dissipation. The setup we realize is also the first experimental example of ``tabletop reversibility'': for a good range of choices of the reference state, the Petz recovery map can be implemented with the same devices as the forward dissipative evolution, whose effect it is partially undoing. Our results demonstrate that the Petz recovery map can be resource-efficiently realized without requiring complex ancillary resources, providing a feasible pathway for mitigating information loss in quantum systems.
    
\end{abstract}

\maketitle

\section{Introduction}
In quantum information theory, the evolution of a closed system is described by a unitary operator acting on its Hilbert space \cite{messiah2014quantum}. Since every unitary operator has an inverse, the information encoded in a closed system can, in principle, always be perfectly recovered. In practice, however, most quantum systems are not perfectly isolated. They inevitably interact with their surrounding environment, leading to open-system dynamics. Such interactions induce environmental noise, such as decoherence and dissipation, and the resulting evolution can no longer be captured by a unitary operator on the system alone \cite{davies1976quantum}, rendering the quantum process irreversible.

The detrimental influence of environmental coupling is one of the challenges in realizing quantum technologies. Decoherence and dissipation rapidly degrade superposition and entanglement, which are the essential resources for tasks such as quantum computation, quantum communication, and quantum sensing \cite{chuang1995quantum,shor1995scheme,van1997ideal,matsuzaki2011magnetic,tsang2013quantum}. To mitigate these effects, various strategies have been developed. In quantum error correction, the relevant information is multiplexed prior to the noisy evolution, so as to be fully recoverable, while irrelevant information is lost on the multiplexed system \cite{shor1995scheme, PhysRevA.32.3266}. In dynamical decoupling, extra controlled pulses are used during the open evolution to mitigate its effect \cite{viola1998dynamical,viola1999dynamical}. When neither prior multiplexing nor active intervention are implemeted, only partial recovery is possible. In this situation, the Petz recovery map has attracted attention as an elegant and general approach, theoretically grounded in quantum information \cite{petz1986sufficient, petz2003monotonicity}. Such a map provides an explicit construction of a recovery channel with near-optimal performance \cite{barnum2002reversing,ng2010simple,zheng2024near,li2025optimality}, offering a reversal procedure based on a chosen reference state, without problem-specific optimization.

The Petz recovery map has been widely explored from a theoretical perspective, including quantum data processing \cite{beigi1504decoding}, fluctuation theorems \cite{aaberg2018fully, kwon2019fluctuation, aw2021fluctuation, buscemi2021fluctuation} and thermodynamical entropies \cite{buscemi2023observational}. 
It is also considered one of the promising candidates for quantum Bayesian retrodiction to define a quantum analog of Bayes' rule \cite{leifer2013towards, tsang2022generalized, parzygnat2023axioms, parzygnat2023time, cenxin2023quantum,Surace2023stateretrieval,  bai2025quantum, liu2025retrodictive, liu2025state, liu2025unifying}. 
Despite its theoretical importance, the experimental realizations of the Petz recovery map are sparse. Recently, concrete proposals and implementations have emerged, including the design of quantum algorithms~\cite{gilyen2022quantum}, experimental demonstrations on ion-trap platforms~\cite{png2025petz} and NMR quantum processors~\cite{singh2025realizing}, and conditions for realizing the map without additional devices~\cite{aw2024role, song2025exact}. While these developments mark important progress, systematic and versatile methods for implementing the Petz recovery map across diverse physical architectures remain largely unexplored.

\begin{figure}[hbtp]
 \centering
 \includegraphics[width=0.5\textwidth]{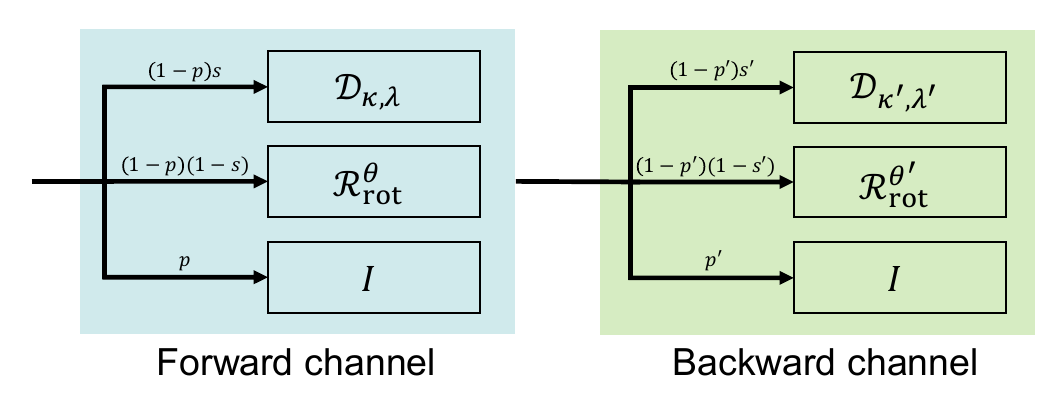}
 \caption{Schematic diagram illustrating the forward channel and the corresponding Petz recovery map acting as the backward channel. Both channels consist of three parts: identity channel $I$, rotation channel $\mathcal{R}_{\text{rot}}$ and the dissipation channel $\mathcal{D}_{\kappa, \lambda}$.}
 \label{Fig. 1_schematic}
 \end{figure}

In this work, we take a further step by implementing the Petz recovery map for a class of qubit channels using a photonic system. We show that for a large class of channels, by selecting appropriate reference states, the Petz recovery map takes the same form as the forward channel, with modified parameters (Fig.~\ref{Fig. 1_schematic}, right). This allows the recovery channel to be implemented with minimal changes to the setup required for the forward dynamics, which saves experimental resources. Our results therefore provide both conceptual insight into the structure of Petz recovery and practical guidance for its realization in physical platforms.

\section{Quantum channel and Petz recovery map}\label{preliminary}
Any quantum dynamics, including those of open quantum systems, can be described by a quantum channel, which is a completely positive and trace-preserving (CPTP) linear map acting on density operators. Formally,  a quantum channel map can be described as $
    \mathcal{E}: \mathcal{S}(\mathcal{H}) \mapsto \mathcal{S}(\mathcal{K})$, 
where $\mathcal{S}(\mathcal{H},\mathcal{K})$ denotes the set of density operators on the Hilbert spaces $\mathcal{H}$ and $\mathcal{K}$, respectively. Quantum channels provide a general framework for describing both unitary evolutions of closed systems and the non-unitary dynamics that arise when a system interacts with an external environment.

Specifically, the action of the channel is expressed as: 
\begin{align}
				\label{operator-sum}
    \mathcal{E}(\rho) = \sum_i K_i \rho K_i^\dagger, 
\end{align}
where the Kraus operators satisfy the completeness relation $\sum_i K_i^\dagger K_i = I$. 

Given the forward channel $\mathcal{E}$, and one set of its Kraus operators $\{ K_i\}_{i \in \mathbb{N}}$, the Petz recovery map $\mathcal{P}$ with reference state $\sigma$ is given by 
\begin{align}
    \label{eq:Petz}
    \begin{aligned}
    \mathcal{P}_{\mathcal{E},\sigma} (\bullet) &= \sqrt{\sigma} \mathcal{E}^\dagger(\frac{1}{\sqrt{\mathcal{E}(\sigma)}} \bullet \frac{1}{\sqrt{\mathcal{E}(\sigma)}}) \sqrt{\sigma}\\
    &= \sum_i \sqrt{\sigma} K_i^\dagger \frac{1}{\sqrt{\mathcal{E}(\sigma)}} \bullet \frac{1}{\sqrt{\mathcal{E}(\sigma)}} K_i \sqrt{\sigma}, 
    \end{aligned}
\end{align}
where $\mathcal{E}^\dagger(\rho) = \sum_i K^{\dagger}_i \rho K_i$ is the adjoint channel defined by $ \Tr (\mathcal{E}(\rho)M) = \Tr (\rho \mathcal{E}^\dagger (M))$. The reference state $\sigma$ serves as a quantum prior and can be perfectly recovered by the Petz recovery map as $\mathcal{P}_{\mathcal{E},\sigma}\mathcal{E}(\sigma) = \sigma$.

We see that the Petz map is written as the sequential composition of three maps: $\mathcal{P}_{\mathcal{E},\sigma}=A_{\sigma}\circ\mathcal{E}^\dagger\circ A_{\mathcal{E}(\sigma)^{-1}}$ where $A_X(\bullet)=\sqrt{X}\bullet \sqrt{X}$. This sequentiality is very handy for calculations, but it should not be used as a blueprint to design an implementation. The reason is that those three maps are not trace preserving in general: thus, a sequential implementation would require postselection. The Petz map is a proper CPTP map, and should be implementable without postselection. This is the path we are going to take here.

 \begin{figure*}[hbtp]
 \centering
 \includegraphics[width=1\textwidth]{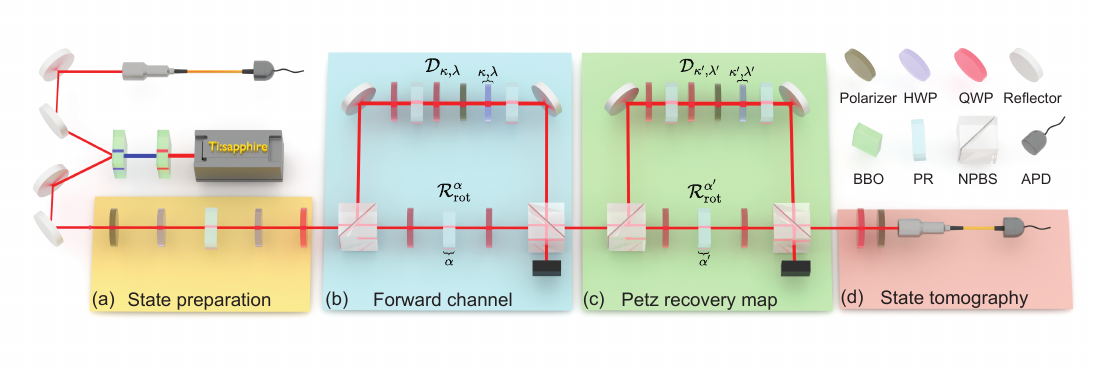}
 \caption{Experimental setup for implementing the forward channel and the Petz recovery map. The setup contains four parts: (a) state preparation, (b) forward channel, (c) Petz recovery map and (d) state tomography. Experimentally, the photons are generated via the spontaneous parametric down-conversion (SPDC) and collected by the APD. The identity part $I$ and the rotation part $\mathcal{R}_{\mathrm{rot}}^{\pi /2}$ ($\mathcal{R}_{\mathrm{rot}}^{\theta}$) of the forward channel (Petz recovery map) are combined and implemented as a single new rotation operation $\mathcal{R}_{\mathrm{rot}}^{\alpha}$ ($\mathcal{R}_{\mathrm{rot}}^{\alpha '}$) (see \cref{Exp. simplify} for detailed derivation). The values of $\kappa^{(')},\lambda^{(')}$ and $\alpha^{(')}$ are determined by adjusting the angle of the corresponding HWP and the thickness of the PR, respectively. BBO, $\beta-$barium borate crystal, HWP half-wave plate, QWP quarter-wave plate, 
 PR phase retarder, NPBS non-polarizing beam splitter, APD Avalanche photodiode.}
 \label{Fig. 2_setup}
 \end{figure*}

\section{Implementation of the Petz recovery map}
\label{implementation}

We investigate the recovery through the Petz map of the initial quantum state after the decoherence and dissipation \sout{loss of information} incurred in the family of quantum channels~\cite{li2025experimental} given by
\begin{align}
    \label{eq:channel}
    \mathcal{E}(\rho) = p\rho + (1-p)[(1-s)\mathcal{R}_{\text{rot}}^{\theta}(\rho)+s\mathcal{D}_{\kappa, \lambda}(\rho)]\,. 
\end{align}
These channels (see \cref{implementation proof} for their Kraus operators) are the incoherent mixture of four channels: the identity channel, with probability $p\in[0,1]$; two unitary channels consisting of rotations around the $y$-axis of the Bloch sphere by an angle either $+\theta$ or $-\theta$, each with probability $(1-p)(1-s)/2$, with $s\in[0,1]$; and a strongly dissipative channel $\mathcal{D}_{\kappa, \lambda}$. This dissipation channel $\mathcal{D}_{\kappa,\lambda}$ removes all coherences and reshuffles the diagonal weights according to $\langle 0|\mathcal{D}_{\kappa, \lambda}(\rho)|0\rangle= \kappa\langle 0|\rho|0\rangle+\lambda \langle 1|\rho|1\rangle$ and $\langle 1|\mathcal{D}_{\kappa, \lambda}(\rho)|1\rangle= (1-\kappa)\langle 0|\rho|0\rangle+(1-\lambda) \langle 1|\rho|1\rangle$, with $\kappa,\lambda\in [0,1]$. Overall, the family of channels $\mathcal{E}$ can lose information through both dephasing and amplitude damping, which is introduced by the dissipative channel $\mathcal{D}_{\kappa,\lambda}$.

For a wide range of the parameters in \cref{eq:channel} and some reference states $\sigma$, the Petz recovery map belongs to the same family of channels,
\begin{align}
    \begin{aligned}
        \label{eq:construction}
        & \mathcal{P}_{\mathcal{E},\sigma}(\rho) \\
        &= p'\rho + (1-p')[(1-s')\mathcal{R}_{\text{rot}}^{\theta'}(\rho)+s'\mathcal{D}_{\kappa', \lambda'}(\rho)], 
    \end{aligned}          
\end{align}
Thus, we can implement the Petz map with the same devices as the forward channel, realizing a specific example of tabletop reversibility \cite{aw2024role, song2025exact}. 

For the following discussion, we choose a diagonal reference state
\begin{align}
    \sigma = r\ketbra{0}{0} + (1-r) \ketbra{1}{1}\,,&\quad r \in [0,1]\,.
\end{align} We have checked numerically that tabletop reversibility holds also for other choices. It can be shown that \cref{eq:construction} determines four conditions on the Choi matrix of the channel, leaving one parameter free: we choose to tune $p'$. In \cref{implementation proof} we give the conditions for tabletop reversibility for a broad range of values of $\theta$, $\kappa$ and $\lambda$. For the main text and the experiment, we set $\theta = \pi /2$, $\kappa = 1$ and $\lambda = 1$; in this case, $\mathcal{D}_{\kappa=\lambda=1}$ is a complete amplitude damping channel that relaxes any input states to the ground state $\vert 0 \rangle$. 

For these choices of parameters, the following Result describes the condition for tabletop reversibility, and the resulting values of the parameters:
\begin{rst}
    \label{result1}
    Let $\mathcal{E}$  be the forward channel of \cref{eq:channel}, with parameters $p, s, \theta = \pi /2, \kappa = 1, \lambda = 1$. Its Petz recovery map is implementable with same structure \cref{eq:construction} if the reference state $\sigma = r\ketbra{0}{0} + (1-r) \ketbra{1}{1}$ is such that 
\begin{align}
    \frac{1}{2}-\frac{1-p}{2(1+p)}s \leq r \leq \frac{1+s}{2}.\label{eq:cond}
\end{align}
When this holds, the Petz map is given by \cref{eq:construction} with
\begin{align}
    \begin{aligned}
        0 &\leq p' < \frac{\sqrt{(1-r)r}(1+p-s+ps)}{\sqrt{1-(p-2pr+ps-s)^2}}, \label{eq:pprime}\\
        \theta' &= 2 \arctan\left( \sqrt{ \frac{1}{2}\frac{ \gamma  (1-p)(1-s)  }{ \gamma \left[ p + (1-p)(1-s) \right] - p' } } \right), \\
        s' &= \frac{ 1 - \gamma \left[ p + (1-p)(1-s) \right] }{ 1 - p' }, \\
        \kappa' &= \frac{1}{2}\frac{ (\frac{r}{q_0}-\gamma) \left[2p + (1-p)(1-s) \right] }{ 1 - \gamma \left[ p + (1-p)(1-s) \right] }, \\
        \lambda' &= \frac{1}{2}\frac{ \left( \frac{r}{q_1} - \gamma \right) (1-p)(1-s)  }{ 1 - \gamma \left[ p + (1-p)(1-s) \right] },
    \end{aligned}
\end{align}
    where
    \begin{align*}
        q_0 &= rp + \frac{1}{2}(1-p)(1+s), \\
        q_1 &= (1-r)p + \frac{1}{2}(1-p)(1-s),\\
        \gamma &= \frac{(1-r)r}{q_0 q_1}. 
    \end{align*}
\end{rst}
The proof is given in \cref{example}. Interestingly, the maximum value $r=\frac{1+s}{2}$ is attained when $\sigma$ is the steady state of the original channel, i.e.~$\mathcal{E}(\sigma)=\sigma$; and in this case, the Petz recovery map is identical to the forward channel (see \cref{sec:steady}). The steady state is often considered a canonical choice of the reference state~\cite{lautenbacher2022approximating,kwon2022reversing}. 

 \begin{figure*}[hbtp]
 \centering
 \includegraphics[width=1\textwidth]{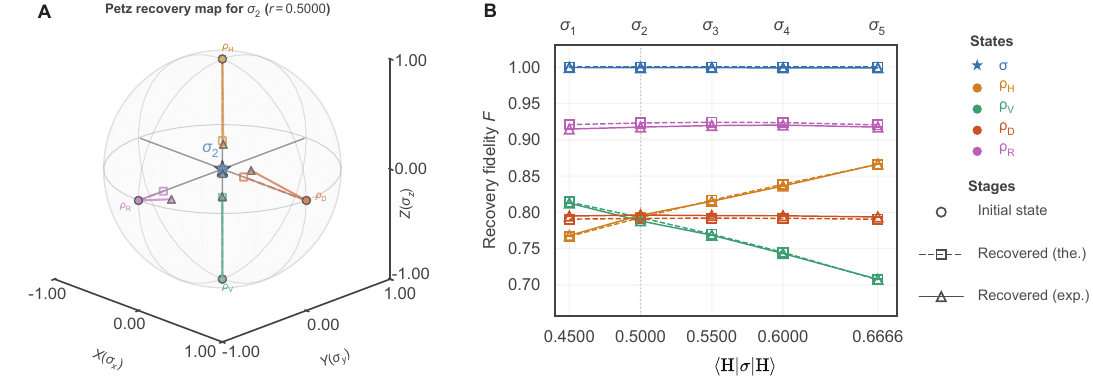}
 \caption{Experimental performance of the Petz recovery map. (A) Recovery trajectories on the Bloch sphere for the reference state $\sigma_2$. Only $\sigma_2$ is shown here, for simplicity, as maps corresponding to other reference states ($\sigma_1,\sigma_3,\sigma_4,\sigma_5$) exhibit similar trajectories. (B) Recovery fidelity $F$ of various input states across different Petz recovery map for different reference states. In both panels, circles with different colors distinguish the input states ($ \rho_{\mathrm{H}}, \rho_{\mathrm{V}}, \rho_{\mathrm{D}}, \rho_{\mathrm{R}}$) with the star specifically indicating the reference state. Squares with dashed lines and triangles with solid lines represent the theoretical and experimental recovered states, respectively. Error bars in panel B are smaller than the data markers.}
\label{Fig. 3_results}
\end{figure*}

\section{Experimental details and results}\label{experiment}
In this section, we present our experimental settings of the forward channel and its Petz recovery map. The qubit is encoded in the polarization degree of freedom of a single photon generated via spontaneous parametric down-conversion, with the horizontal and vertical polarization states corresponding to $\ket{0}$ and $\ket{1}$, respectively. As shown in \cref{Fig. 2_setup}, the experimental setup consists of four modules: state preparation, the forward channel, the Petz recovery map, and state tomography.

At the state preparation stage, the input state $\rho$ is prepared by sequentially passing through a half-wave plate (HWP), phase retarders (PRs), and a quarter-wave plate (QWP). The PRs are implemented using liquid-crystal variable retarders (LCVR) to introduce precise mixing of different phases, thereby controlling the degree of mixture in the prepared state. In the experiment, the forward channel parameters are fixed at $p = 1/2, s = 1/3$ and the parameters are chosen as $p'= 1/2$ in the Petz recovery map. Constrained by \cref{result1}, the implementable reference states should satisfy $4/9 \leq r \leq 2/3$. Within this range, we prepare five distinct reference states $\sigma = r|0\rangle\langle 0\vert + \left( 1-r \right)|1\rangle \langle 1\vert$ with $r = \{0.4500, 0.5000, 0.5500, 0.6000, 0.6666\}$. The reference state can be prepared by using a HWP and a PR. To quantify the state preparation quality, we experimentally perform the tomography of the prepared states utilizing the state fidelity \cite{uhlmann1976fidelity, jozsa1994fidelity}, defined as $F(\rho_1, \rho_2) = (\Tr \sqrt{\sqrt{\rho_2}\rho_1 \sqrt{\rho_2}})^2$. The fidelities between the experimentally prepared reference state and the theoretical values are $\mathcal{F}_{\sigma_{1}} = 0.9973 \pm 0.0008,\mathcal{F}_{\sigma_{2}} = 0.9977\pm 0.0007,   \mathcal{F}_{\sigma_{3}} = 0.9991 \pm 0.0006,   \mathcal{F}_{\sigma_{4}} = 0.9984 \pm 0.0008, \mathcal{F}_{\sigma_{5}} = 0.9982 \pm 0.0008,$ respectively. 

For the fixed forward channel, we implement five corresponding Petz recovery maps based on five distinct reference states $\sigma$. The forward channel and Petz recovery map are implemented using two non-polarizing beam splitters (NPBS), several wave plates (WPs) and PRs. For both channels, the first NPBS divides the light into the transmitted path and the reflected path. The transmitted path combines the identity part $\rho$ and the rotation part $\mathcal{R}^{\pi /2}_{\text{rot}}$ into a new rotation part $\mathcal{R}^{\alpha}_{\mathrm{\mathrm{rot}}}$ with a new rotation angle $\alpha$ (see \cref{Exp. simplify} for a detailed derivation), while the reflected path achieves the dissipation part $\mathcal{D}_{\kappa=\lambda = 1}$. The experimental layout of the Petz recovery map shares the same structure as the forward channel but with modified parameters $\mathcal{R}_{\mathrm{rot}}^{\alpha'}$ and $\mathcal{D}_{\kappa',\lambda'}$, realized by fine-tuning the HWP angles and the PR thickness, respectively. Notably, all the decoherence processes (both complete and partial) throughout the entire experimental setup are realized via mixing different phases using PRs (see \cref{Exp. detail} for rigorous derivations).

The state tomography is performed using a QWP followed by a polarizer, and the photons are then detected by avalanche photodiode (APD) single-photon detectors. With this setup, we comprehensively reconstruct the density matrix of the initial input states, the five distinct reference states and the final evolved states. Theoretically, a Petz recovery map perfectly recovers its specific reference state from the noisy channel, i.e., $\mathcal{P}_{\mathcal{E},\sigma}(\mathcal{E}(\sigma))=\sigma$. Relying on this fundamental property, we demonstrate that the channels we implement experimentally are the desired Petz recovery maps with high precision.
The fidelities between the experimentally recovered reference state and the theoretical reference state are  $\mathcal{F}_{\mathcal{P}\mathcal{E}({{\sigma_{1}}})} = 0.9995 \pm 0.0001, \mathcal{F}_{\mathcal{P}\mathcal{E}({{\sigma_{2}}})} = 0.9992 \pm 0.0002,   \mathcal{F}_{\mathcal{P}\mathcal{E}({{\sigma_{3}}})} = 0.9993 \pm 0.0004, \mathcal{F}_{\mathcal{P}\mathcal{E}({{\sigma_{4}}})} = 0.9991 \pm 0.0002,  \mathcal{F}_{\mathcal{P}\mathcal{E}({\sigma_{5}})} = 0.9990 \pm 0.0002$. 
As shown in Fig. \ref{Fig. 3_results}, the input reference state (the blue star) and the theoretical recovery reference state (the blue rectangle) are in the same position, while the experimentally recovered reference states (blue triangles) are close to them.

To evaluate the extent to which the non-reference state can be recovered by the Petz recovery map, we prepare four typical qubit states as inputs: $\rho_{\text{H}} = \ket{\mathrm{H}}\bra{\mathrm{H}}, \rho_{\text{V}} = \ket{\mathrm{V}}\bra{\mathrm{V}}, \rho_{\text{D}} = \ket{\mathrm{D}}\bra{\mathrm{D}}, \rho_{\text{R}} = \ket{\mathrm{R}}\bra{\mathrm{R}}$, where $\ket{\mathrm{D}} = (\ket{\mathrm{H}} + \ket{\mathrm{V}})/\sqrt{2}$ and $\ket{\mathrm{R}} = (\ket{\mathrm{H}} -i\ket{\mathrm{V}})/\sqrt{2}$. The fidelities between the experimentally prepared input state and the theoretical states are $\mathcal{F}_{\rho_{\text{H}}} = 0.9999 \pm 0.00006, \mathcal{F}_{\rho_{\text{V}}} = 0.9986 \pm 0.0003, \mathcal{F}_{\rho_{\text{D}}} = 0.9991 \pm 0.0003, \mathcal{F}_{\rho_{\text{R}}} = 0.9986 \pm 0.0002 $ respectively. The experimentally recovered states for the different input states and the five different Petz recovery maps are shown in Fig. \ref{Fig. 3_results}. We can see that the recovery performance depends on how well the reference state matches the diagonal structure of the input state. In particular, $\rho_{\text{H}}$ is closest to $\sigma_{5}$, and $\rho_{\text{V}}$ is closest to $\sigma_{1}$. They exhibit better recovery performance with the corresponding Petz recovery map. As the diagonal elements of $\rho_{\text{D}}$ and $\rho_{\text{R}}$ coincide with those of  the diagonal elements of $\sigma_{2}$; the corresponding Petz recovery map with $\sigma_{2}$ restores their diagonal terms particularly well. As a result, for these specific equatorial input states, their recovered counterparts under $\sigma_2$ remain geometrically confined close to the equator of the Bloch sphere.

\begin{table*}[htbp]
\centering
\caption{\textbf{Comparison of Fidelity and Trace Distance across different parameters.} 
The values highlighted in bold correspond to the optimal or selected data points as indicated in the original analysis.}
\label{tab:results}
\resizebox{\textwidth}{!}{
\begin{tabular}{c c c c c c c c c c c}
\toprule
\multirow{2}{*}{$\sigma$} &\multicolumn{2}{c}{ $\mathcal{\mathcal{P}\mathcal{E}}(\sigma)$} & \multicolumn{2}{c}{ $\mathcal{\mathcal{P}\mathcal{E}}(\rho_{\text{H}})$} & \multicolumn{2}{c}{$\mathcal{P}\mathcal{E}(\rho_{\text{V}})$} & \multicolumn{2}{c}{$\mathcal{\mathcal{P}\mathcal{E}(\rho_{\mathrm{D}})}$ } & \multicolumn{2}{c}{$\mathcal{\mathcal{P}\mathcal{E}}(\rho_{\mathrm{R}})$} \\
\cmidrule(lr){2-3} \cmidrule(lr){4-5} \cmidrule(lr){6-7} \cmidrule(lr){8-9}\cmidrule(lr){10-11}
  & Fidelity & Trace Dist. & Fidelity & Trace Dist. & Fidelity & Trace Dist. & Fidelity & Trace Dist. & Fidelity & Trace Dist. \\
\midrule
$\sigma_{1}$ &0.9995&0.0317& 0.7668& 0.4120   &0.8142 & 0.3371&  0.7907 & 0.3767 & 0.9208 & 0.1567 \\
$\sigma_{2}$ &0.9992&0.0407&0.7928& 0.3714   &0.7928 & 0.3714&  0.7917 & 0.3732 & 0.9231 & 0.1479 \\
$\sigma_{3}$ &0.9993&0.0383& 0.8166& 0.3331   &0.7699 & 0.4072&  0.7921 & 0.3744 & 0.9241 & 0.1507 \\
$\sigma_{4}$ &0.9991&0.0434&0.8386& 0.2967   &0.7450 & 0.4450&  0.7918 & 0.3804 & 0.9236 & 0.1647 \\
$\sigma_{5}$ &0.9990&0.0441& 0.8660& 0.2500   &0.7071 & 0.5000&  0.7906 & 0.3953 & 0.9204 & 0.1974 \\
\bottomrule
\end{tabular}%
}
\end{table*}

Crucially, when employing a diagonal reference state, the Petz recovery map predominantly restores the populations (diagonal elements) of the input state, leaving the coherences (off-diagonal elements) of the state after the forward channel largely unchanged. As a result,  fidelity becomes an inadequate metric for equatorial states on the Bloch sphere, such as $|\mathrm{D}\rangle$ 
and $|\mathrm{R}\rangle$). For these states, the fidelity is largely insensitive to the diagonal variations and is heavily dominated by residual off-diagonal coherence.
To overcome this limitation, we therefore employ the trace distance \cite{nielsen2010quantum}, defined as  and $T(\rho_1,\rho_2) = [ \Tr \sqrt{(\rho_1-\rho_2)^\dagger (\rho_1-\rho_2) }]/2$, which offers enhanced sensitivity to the overall deviation between the recovered states and the input states. Combining both fidelity and trace distance thereby yields a rigorously comprehensive evaluation of the recovery performance. The full set of these two metrics for experimentally recovered states and the theoretical states is summarized in \cref{tab:results}.

\section{Conclusions}
In this work, we have theoretically derived and experimentally realized the Petz recovery map for a family of qubit channels within a quantum photonic system. We show that by selecting an appropriate diagonal reference state, the Petz recovery map can be implemented using the identical structural form as the forward channel, requiring only a modification of the experimental parameters. This approach offers a significant advantage in resource efficiency, as it allows the recovery channel to be constructed with minimal alterations to the setup used for the forward dynamics.

Experimentally, we validate this framework using a linear optical setup, verifying the recovery performance through quantum state tomography. The experimental results confirm that the Petz recovery map can be effectively realized when the input state deviates from the ideal reference. 

In summary, our results provide a practical guidance for the implementation of Petz recovery map. By demonstrating that a complex recovery map can be realized through a resource-efficient reconfiguration of the forward channel, this work shares one practical way for realizing Petz recovery across different physical platforms.

\begin{acknowledgments}
We thank Clive Aw for the helpful discussions. Hui Li, Yue Pan, Liang Xu and Lijian Zhang are supported by the Quantum Science and Technology-National Science and Technology Major Project (Grant No. 2024ZD0300900), National Key Research and Development Program of China (Grant No. 2023YFC2205802), National Natural Science Foundation of China (Grant Nos. U24A2017, 12347104, 12461160276, 12504418 and 92576111), Natural Science Foundation of Jiangsu Province (Grant Nos. BK20243060, BK20233001 and BK20251994), Fundamental and Interdisciplinary Disciplines Breakthrough Plan of the Ministry of Education of China (Grant No. JYB2025XDXM105). 
Jinyan Chen, Minjeong Song and Valerio Scarani are supported by the National Research Foundation, Singapore through the National Quantum Office, hosted in A*STAR, under its Centre for Quantum Technologies Funding Initiative (S24Q2d0009); and by the Ministry of Education, Singapore, under the Tier 2 grant ``Bayesian approach to irreversibility'' (Grant No.~MOE-T2EP50123-0002). 
\end{acknowledgments}

\bibliography{reference}

@book{messiah2014quantum,
  title={Quantum mechanics},
  author={Messiah, Albert},
  year={2014},
  publisher={Courier Corporation}
}

@book{davies1976quantum,
  title={Quantum Theory of Open Systems},
  author={Davies, E.B.},
  isbn={9780122061509},
  lccn={76016963},
  url={https://books.google.com.sg/books?id=I5kuAAAAIAAJ},
  year={1976},
  publisher={Academic Press}
}

@article{chuang1995quantum,
  title={Quantum computers, factoring, and decoherence},
  author={Chuang, Isaac L and Laflamme, Raymond and Shor, Peter W and Zurek, Wojciech H},
  journal={Science},
  volume={270},
  number={5242},
  pages={1633--1635},
  year={1995},
  publisher={American Association for the Advancement of Science},
  doi = {10.1126/science.270.5242.1633}
}

@article{shor1995scheme,
  title={Scheme for reducing decoherence in quantum computer memory},
  author={Shor, Peter W},
  journal={Physical Review A},
  volume={52},
  number={4},
  pages={R2493},
  year={1995},
  publisher={APS},
  doi = {https://doi.org/10.1103/PhysRevA.52.R2493}
}

@article{van1997ideal,
  title={Ideal quantum communication over noisy channels: a quantum optical implementation},
  author={van Enk, Steven J and Cirac, Juan I and Zoller, Peter},
  journal={Physical Review Letters},
  volume={78},
  number={22},
  pages={4293},
  year={1997},
  publisher={APS},
  doi = {https://doi.org/10.1103/PhysRevLett.78.4293}
}

@article{matsuzaki2011magnetic,
  title={Magnetic field sensing beyond the standard quantum limit under the effect of decoherence},
  author={Matsuzaki, Yuichiro and Benjamin, Simon C and Fitzsimons, Joseph},
  journal={Physical Review A—Atomic, Molecular, and Optical Physics},
  volume={84},
  number={1},
  pages={012103},
  year={2011},
  publisher={APS},
  doi = {https://doi.org/10.1103/PhysRevA.84.012103}
}

@article{tsang2013quantum,
  title={Quantum metrology with open dynamical systems},
  author={Tsang, Mankei},
  journal={New Journal of Physics},
  volume={15},
  number={7},
  pages={073005},
  year={2013},
  publisher={IOP Publishing},
  doi = {10.1088/1367-2630/15/7/073005}
}

@article{PhysRevA.32.3266,
  title = {Reversible logic and quantum computers},
  author = {Peres, Asher},
  journal = {Phys. Rev. A},
  volume = {32},
  issue = {6},
  pages = {3266--3276},
  numpages = {0},
  year = {1985},
  month = {Dec},
  publisher = {American Physical Society},
  doi = {10.1103/PhysRevA.32.3266},
  url = {https://link.aps.org/doi/10.1103/PhysRevA.32.3266}
}

@article{viola1998dynamical,
  title={Dynamical suppression of decoherence in two-state quantum systems},
  author={Viola, Lorenza and Lloyd, Seth},
  journal={Physical Review A},
  volume={58},
  number={4},
  pages={2733},
  year={1998},
  publisher={APS},
  doi = {https://doi.org/10.1103/PhysRevA.58.2733}
}

@article{viola1999dynamical,
  title={Dynamical decoupling of open quantum systems},
  author={Viola, Lorenza and Knill, Emanuel and Lloyd, Seth},
  journal={Physical Review Letters},
  volume={82},
  number={12},
  pages={2417},
  year={1999},
  publisher={APS},
  doi = {https://doi.org/10.1103/PhysRevLett.82.2417}
}

@article{petz1986sufficient,
  title={Sufficient subalgebras and the relative entropy of states of a von {N}eumann algebra},
  author={Petz, D{\'e}nes},
  journal={Communications in mathematical physics},
  volume={105},
  number={1},
  pages={123--131},
  year={1986},
  publisher={Springer},
  doi = {10.1007/bf01212345}
}

@article{petz2003monotonicity,
  title={Monotonicity of quantum relative entropy revisited},
  author={Petz, D{\'e}nes},
  journal={Reviews in Mathematical Physics},
  volume={15},
  number={01},
  pages={79--91},
  year={2003},
  publisher={World Scientific},
  doi = {https://doi.org/10.1142/S0129055X03001576}
}

@article{barnum2002reversing,
  title={Reversing quantum dynamics with near-optimal quantum and classical fidelity},
  author={Barnum, Howard and Knill, Emanuel},
  journal={Journal of Mathematical Physics},
  volume={43},
  number={5},
  pages={2097--2106},
  year={2002},
  publisher={American Institute of Physics},
  doi = {https://doi.org/10.1063/1.1459754}
}

@article{ng2010simple,
  title={Simple approach to approximate quantum error correction based on the transpose channel},
  author={Ng, Hui Khoon and Mandayam, Prabha},
  journal={Physical Review A—Atomic, Molecular, and Optical Physics},
  volume={81},
  number={6},
  pages={062342},
  year={2010},
  publisher={APS},
  doi = {https://doi.org/10.1103/PhysRevA.81.062342}
}

@article{zheng2024near,
  title={Near-optimal performance of quantum error correction codes},
  author={Zheng, Guo and He, Wenhao and Lee, Gideon and Jiang, Liang},
  journal={Physical Review Letters},
  volume={132},
  number={25},
  pages={250602},
  year={2024},
  publisher={APS},
  doi = {https://doi.org/10.1103/PhysRevLett.132.250602}
}

@article{li2025optimality,
  title={Optimality Condition for the {P}etz Map},
  author={Li, Bikun and Wang, Zhaoyou and Zheng, Guo and Wong, Yat and Jiang, Liang},
  journal={Physical Review Letters},
  volume={134},
  number={20},
  pages={200602},
  year={2025},
  publisher={APS},
  doi = {https://doi.org/10.1103/PhysRevLett.134.200602}
}

@article{beigi1504decoding,
  title={Decoding quantum information via the {P}etz recovery map},
  author={Beigi, Salman and Datta, Nilanjana and Leditzky, Felix},
  journal={arXiv preprint arXiv:1504.04449},
  year={2015},
  doi = {https://doi.org/10.1063/1.4961515}
}

@article{aaberg2018fully,
  title={Fully quantum fluctuation theorems},
  author={{\AA}berg, Johan},
  journal={Physical Review X},
  volume={8},
  number={1},
  pages={011019},
  year={2018},
  publisher={APS},
  doi = {https://doi.org/10.1103/PhysRevX.8.011019}
}

@article{kwon2019fluctuation,
  title={Fluctuation theorems for a quantum channel},
  author={Kwon, Hyukjoon and Kim, MS},
  journal={Physical Review X},
  volume={9},
  number={3},
  pages={031029},
  year={2019},
  publisher={APS},
  doi = {https://doi.org/10.1103/PhysRevX.9.031029}
}

@article{aw2021fluctuation,
  title={Fluctuation theorems with retrodiction rather than reverse processes},
  author={Aw, Clive Cenxin and Buscemi, Francesco and Scarani, Valerio},
  journal={AVS Quantum Science},
  volume={3},
  number={4},
  year={2021},
  publisher={AIP Publishing},
  doi = {https://doi.org/10.1116/5.0060893}
}

@article{buscemi2021fluctuation,
  title={Fluctuation theorems from {B}ayesian retrodiction},
  author={Buscemi, Francesco and Scarani, Valerio},
  journal={Physical Review E},
  volume={103},
  number={5},
  pages={052111},
  year={2021},
  publisher={APS}, 
  doi = {https://doi.org/10.1103/PhysRevE.103.052111}
}

@article{buscemi2023observational,
  title={Observational entropy, coarse-grained states, and the {P}etz recovery map: information-theoretic properties and bounds},
  author={Buscemi, Francesco and Schindler, Joseph and {\v{S}}afr{\'a}nek, Dominik},
  journal={New Journal of Physics},
  volume={25},
  number={5},
  pages={053002},
  year={2023},
  publisher={IOP Publishing},
  doi = {10.1088/1367-2630/accd11}
}

@article{leifer2013towards,
  title={Towards a formulation of quantum theory as a causally neutral theory of {B}ayesian inference},
  author={Leifer, Matthew S and Spekkens, Robert W},
  journal={Physical Review A—Atomic, Molecular, and Optical Physics},
  volume={88},
  number={5},
  pages={052130},
  year={2013},
  publisher={APS},
  doi = {https://doi.org/10.1103/PhysRevA.88.052130}
}

@article{tsang2022generalized,
  title={Generalized conditional expectations for quantum retrodiction and smoothing},
  author={Tsang, Mankei},
  journal={Physical Review A},
  volume={105},
  number={4},
  pages={042213},
  year={2022},
  publisher={APS},
  doi = {https://doi.org/10.1103/PhysRevA.105.042213}
}

@article{parzygnat2023axioms,
  title={Axioms for retrodiction: achieving time-reversal symmetry with a prior},
  author={Parzygnat, Arthur J and Buscemi, Francesco},
  journal={Quantum},
  volume={7},
  pages={1013},
  year={2023},
  publisher={Verein zur F{\"o}rderung des Open Access Publizierens in den Quantenwissenschaften},
  doi = {https://doi.org/10.22331/q-2023-05-23-1013}
}

@article{parzygnat2023time,
  title={From time-reversal symmetry to quantum {B}ayes’ rules},
  author={Parzygnat, Arthur J and Fullwood, James},
  journal={PRX Quantum},
  volume={4},
  number={2},
  pages={020334},
  year={2023},
  publisher={APS},
  doi = {https://doi.org/10.1103/PRXQuantum.4.020334}
}

@article{cenxin2023quantum,
  title={Quantum {B}ayesian inference in quasiprobability representations},
  author={Cenxin, Aw Clive and Onggadinata, Kelvin and Kaszlikowski, Dagomir and Scarani, Valerio},
  journal={PRX Quantum},
  volume={4},
  number={2},
  pages={020352},
  year={2023},
  publisher={APS},
  doi = {https://doi.org/10.1103/PRXQuantum.4.020352}
}

@article{bai2025quantum,
  title={Quantum {B}ayes’ rule and {P}etz transpose map from the minimum change principle},
  author={Bai, Ge and Buscemi, Francesco and Scarani, Valerio},
  journal={Physical Review Letters},
  volume={135},
  number={9},
  pages={090203},
  year={2025},
  publisher={APS},
  doi = {https://doi.org/10.1103/5n4p-bxhm}
}

@article{liu2025retrodictive,
  title={Retrodictive approach to quantum state smoothing},
  author={Liu, Mingxuan and Scarani, Valerio and Auff{\`e}ves, Alexia and Laverick, Kiarn T},
  journal={Physical Review A},
  volume={112},
  number={3},
  pages={L030203},
  year={2025},
  publisher={APS},
  doi = {https://doi.org/10.1103/8pc3-7pg5}
}

@article{liu2025state,
  title={The state of a quantum system is not a complete description for retrodiction},
  author={Liu, Mingxuan and Bai, Ge and Scarani, Valerio},
  journal={arXiv preprint arXiv:2502.10030},
  year={2025},
  doi = {https://doi.org/10.48550/arXiv.2502.10030}
}

@article{liu2025unifying,
  title={Unifying Quantum Smoothing Theories with Extended Retrodiction},
  author={Liu, Mingxuan and Bai, Ge and Scarani, Valerio},
  journal={arXiv preprint arXiv:2510.08447},
  year={2025},
  doi = {https://doi.org/10.48550/arXiv.2510.08447}
}

@article{gilyen2022quantum,
  title={Quantum algorithm for {P}etz recovery channels and pretty good measurements},
  author={Gily{\'e}n, Andr{\'a}s and Lloyd, Seth and Marvian, Iman and Quek, Yihui and Wilde, Mark M},
  journal={Physical Review Letters},
  volume={128},
  number={22},
  pages={220502},
  year={2022},
  publisher={APS},
  doi = {https://doi.org/10.1103/PhysRevLett.128.220502}
}

@article{png2025petz,
  title={Petz recovery maps of single-qubit decoherence channels in an ion trap quantum processor},
  author={Png, Wen-Han and Scarani, Valerio},
  journal={Physical Review A},
  volume={112},
  number={2},
  pages={022613},
  year={2025},
  publisher={APS},
  doi = {https://doi.org/10.1103/7f8x-n2np}
}

@article{singh2025realizing,
  title={Realizing the {P}etz Recovery Map on an {NMR} Quantum Processor},
  author={Singh, Gayatri and Sahani, Ram Sagar and Jagadish, Vinayak and Lautenbacher, Lea and Bernardes, Nadja K and Dorai, Kavita},
  journal={arXiv preprint arXiv:2508.08998},
  year={2025},
  doi = {https://doi.org/10.48550/arXiv.2508.08998}
}

@article{aw2024role,
  title={Role of dilations in reversing physical processes: Tabletop reversibility and generalized thermal operations},
  author={Aw, Clive Cenxin and Zaw, Lin Htoo and Balanz{\'o}-Juand{\'o}, Maria and Scarani, Valerio},
  journal={PRX Quantum},
  volume={5},
  number={1},
  pages={010332},
  year={2024},
  publisher={APS}
}

@article{song2025exact,
  title={Exact and approximate conditions of tabletop reversibility: when is {P}etz recovery cost-free?},
  author={Song, Minjeong and Kwon, Hyukjoon and Scarani, Valerio},
  journal={arXiv preprint arXiv:2510.26895},
  year={2025},
  doi = {https://doi.org/10.48550/arXiv.2510.26895}
}

@article{li2025experimental,
  title={Experimental demonstration of generalized quantum fluctuation theorems in the presence of coherence},
  author={Li, Hui and Xie, Jie and Kwon, Hyukjoon and Zhao, Yixin and Kim, MS and Zhang, Lijian},
  journal={Science Advances},
  volume={11},
  number={22},
  pages={eadq6014},
  year={2025},
  publisher={American Association for the Advancement of Science}
}

@article{jozsa1994fidelity,
  title={Fidelity for mixed quantum states},
  author={Jozsa, Richard},
  journal={Journal of modern optics},
  volume={41},
  number={12},
  pages={2315--2323},
  year={1994},
  publisher={Taylor \& Francis},
  doi = {https://doi.org/10.1080/09500349414552171}
}

@article{uhlmann1976fidelity,
  title={The ``transition probability'' in the state space of a*-algebra},
  author={Uhlmann, Armin},
  journal={Reports on Mathematical Physics},
  volume={9},
  number={2},
  pages={273--279},
  year={1976},
  publisher={Elsevier},
  doi = {https://doi.org/10.1016/0034-4877(76)90060-4}
}

@book{nielsen2010quantum,
  title={Quantum computation and quantum information},
  author={Nielsen, Michael A and Chuang, Isaac L},
  year={2010},
  publisher={Cambridge university press}
}

@article{kwon2022reversing,
  title={Reversing {L}indblad dynamics via continuous {P}etz recovery map},
  author={Kwon, Hyukjoon and Mukherjee, Rick and Kim, MS},
  journal={Physical Review Letters},
  volume={128},
  number={2},
  pages={020403},
  year={2022},
  publisher={APS}
}

@article{lautenbacher2022approximating,
  title={Approximating invertible maps by recovery channels: Optimality and an application to non-{M}arkovian dynamics},
  author={Lautenbacher, Lea and de Melo, Fernando and Bernardes, Nadja K},
  journal={Physical Review A},
  volume={105},
  number={4},
  pages={042421},
  year={2022},
  publisher={APS}
}

@article{Surace2023stateretrieval,
  doi = {10.22331/q-2023-04-27-990},
  url = {https://doi.org/10.22331/q-2023-04-27-990},
  title = {State retrieval beyond {B}ayes' retrodiction},
  author = {Surace, Jacopo and Scandi, Matteo},
  journal = {{Quantum}},
  issn = {2521-327X},
  publisher = {{Verein zur F{\"{o}}rderung des Open Access Publizierens in den Quantenwissenschaften}},
  volume = {7},
  pages = {990},
  month = apr,
  year = {2023}
}

\appendix
\widetext
\setcounter{section}{0}
\section{\label{implementation proof}Implementation of Petz recovery map}

Here we prove the claim that, in some cases, the Petz map takes the same form as the forward channel, possibly with different parameters.

Consider the forward channel $\mathcal{E}(\rho)$ defined as
\begin{align}
\mathcal{E}(\rho) = p\rho + (1-p)[(1-s)\mathcal{R}_{\mathrm{rot}}^{\theta}(\rho)+s\mathcal{D}_{\kappa, \lambda}(\rho)],\label{channel}
\end{align}
with the corresponding Kraus operators $\{K_i\}$ given by:
\begin{align}
    \begin{split}
        K_0 &= \sqrt{p}\begin{pmatrix} 1 & 0 \\ 0 & 1 \end{pmatrix}, \\
        K_1 &= \sqrt{\frac{(1-p)(1-s)}{2}}\begin{pmatrix} \cos \frac{\theta}{2} & \sin \frac{\theta}{2} \\ -\sin \frac{\theta}{2} & \cos \frac{\theta}{2} \end{pmatrix}, 
        K_2 = \sqrt{\frac{(1-p)(1-s)}{2}}\begin{pmatrix} \cos \frac{\theta}{2} & -\sin \frac{\theta}{2} \\ \sin \frac{\theta}{2} & \cos \frac{\theta}{2} \end{pmatrix}, \\
        K_3 &= \sqrt{(1-p)s\kappa}\begin{pmatrix} 1 & 0 \\ 0 & 0 \end{pmatrix}, 
        K_4 = \sqrt{(1-p)s(1-\kappa)}\begin{pmatrix} 0 & 0 \\ 1 & 0 \end{pmatrix}, \\
        K_5 &= \sqrt{(1-p)s\lambda}\begin{pmatrix} 0 & 1 \\ 0 & 0 \end{pmatrix}, 
        K_6 = \sqrt{(1-p)s(1-\lambda)}\begin{pmatrix} 0 & 0 \\ 0 & 1 \end{pmatrix}. 
    \end{split}
\end{align}
We aim to determine if the Petz recovery map relative to a diagonal reference state \begin{align}
    \sigma = \begin{pmatrix} r & 0 \\ 0 & 1-r \end{pmatrix} \label{sdiag}
\end{align}
can be a channel of the same class \eqref{channel} with a modified set of parameters $p', s', \theta', \kappa', \lambda'$. The formal solution will be given in \cref{eq:general solution} with $p'$ a free parameter, and is valid whenever the parameters are in the 

The derivation in the following relies on the fact that if two channels $\mathcal{E}$ and $\mathcal{E'}$ are equivalent, their Choi matrices are equal, i.e., $J(\mathcal{E}) = J(\mathcal{E'})$, where the Choi matrix is defined as:
\begin{align}
    J(\mathcal{E}) =  (\mathbb{I} \otimes \mathcal{E}) \sum_{i,j}( |i\rangle\langle j| \otimes |i\rangle\langle j|) = \sum_{i,j} |i\rangle\langle j| \otimes \mathcal{E}(|i\rangle\langle j|). 
\end{align}
Denoting the vectorized Kraus operators as the bipartite column vectors $|K_i\rangle \! \rangle = (\mathbb{I} \otimes K_i) \sum_j |j\rangle\otimes |j\rangle$, the Choi matrix simplifies to:
\begin{align}
    J(\mathcal{E}) = \sum_i |K_i\rangle \!\rangle \langle \! \langle K_i|, \label{eq:choi_sum}
\end{align}
where $\{K_i\}$ are the Kraus operators of the channel $\mathcal{E}$. 

We begin by calculating the Choi matrix $J(\mathcal{E})$. Defining the standard ordered computational basis for the bipartite space as $\{|00\rangle, |01\rangle, |10\rangle, |11\rangle\}$, the vectorized form of a general $2\times2$ matrix $A = \begin{pmatrix} a & b \\ c & d \end{pmatrix}$ is $|A\rangle\!\rangle = (a, c, b, d)^T$. Applying this to the Kraus operators $\{K_0, \dots, K_6\}$, we obtain the column vectors:
\begin{align}
    \begin{split}
        |K_0\rangle\!\rangle &= \sqrt{p} (1, 0, 0, 1)^T, \\
        |K_1\rangle\!\rangle &= \sqrt{\frac{(1-p)(1-s)}{2}} \left(\cos\frac{\theta}{2}, -\sin\frac{\theta}{2}, \sin\frac{\theta}{2}, \cos\frac{\theta}{2}\right)^T, \\
        |K_2\rangle\!\rangle &= \sqrt{\frac{(1-p)(1-s)}{2}} \left(\cos\frac{\theta}{2}, \sin\frac{\theta}{2}, -\sin\frac{\theta}{2}, \cos\frac{\theta}{2}\right)^T, \\
        |K_3\rangle\!\rangle &= \sqrt{(1-p)s\kappa} (1, 0, 0, 0)^T, \\
        |K_4\rangle\!\rangle &= \sqrt{(1-p)s(1-\kappa)} (0, 1, 0, 0)^T, \\
        |K_5\rangle\!\rangle &= \sqrt{(1-p)s\lambda} (0, 0, 1, 0)^T, \\
        |K_6\rangle\!\rangle &= \sqrt{(1-p)s(1-\lambda)} (0, 0, 0, 1)^T.
    \end{split}
\end{align}
Summing the outer products $J(\mathcal{E}) = \sum_{j=0}^6 |K_i\rangle\!\rangle\langle\!\langle K_i|$, the cross-terms in $|K_1\rangle\!\rangle\langle\!\langle K_1| + |K_2\rangle\!\rangle\langle\!\langle K_2|$ naturally cancel, resulting in $J(\mathcal{E})$ of the following form
\begin{align}
    J(\mathcal{E}) = 
    \begin{pmatrix}
    j_{00} & 0 & 0 & j_{03} \\
    0 & j_{11} & j_{12} & 0 \\
    0 & j_{21} & j_{22} & 0 \\
    j_{30} & 0 & 0 & j_{33}
    \end{pmatrix},
\end{align}
where the components are given by:
\begin{align}
    j_{00} &= p + (1-p)(1-s)\cos^2\frac{\theta}{2} + (1-p)s\kappa, \nonumber \\
    j_{11} &= (1-p)(1-s)\sin^2\frac{\theta}{2} + (1-p)s(1-\kappa), \nonumber \\
    j_{22} &= (1-p)(1-s)\sin^2\frac{\theta}{2} + (1-p)s\lambda, \nonumber \\
    j_{33} &= p + (1-p)(1-s)\cos^2\frac{\theta}{2} + (1-p)s(1-\lambda), \nonumber \\
    j_{03} &= j_{30} = p + (1-p)(1-s)\cos^2\frac{\theta}{2}, \nonumber \\
    j_{12} &= j_{21} = -(1-p)(1-s)\sin^2\frac{\theta}{2}.
\end{align}
It is not difficult to find that $j_{00}+ j_{11} = 1, j_{22}+j_{33} = 1$. 

Next, we determine the Choi matrix of the corresponding Petz recovery map. By definition, the Petz recovery map $\mathcal{P}_{\mathcal{E},\sigma}$ associated with a channel $\mathcal{E}$ and a full-rank reference state $\sigma$ acts on an input state $\omega$ as:
\begin{align}
    \mathcal{P}_{\mathcal{E},\sigma}(\omega) = \sigma^{1/2} \mathcal{E}^\dagger \left( \mathcal{E}(\sigma)^{-1/2} \omega \mathcal{E}(\sigma)^{-1/2} \right) \sigma^{1/2},
\end{align}
where $\mathcal{E}^\dagger(X) = \sum_i K_i^\dagger X K_i$ is the adjoint map. It directly follows that the Kraus operators for the Petz recovery map, denoted as $\{P_i\}$, take the form $P_i = \sigma^{1/2} K_i^\dagger \mathcal{E}(\sigma)^{-1/2}$.

The assumption \eqref{sdiag} that the reference state is diagonal, together with the relation $\mathcal{E}(\sigma) = \text{Tr}_\mathcal{H}[J(\mathcal{E})(\sigma \otimes \mathbb{I})]$ for the channel $\mathcal{E}:\mathcal{S}(\mathcal{H}) \to \mathcal{S}(\mathcal{K})$, gives
\begin{align}
    \mathcal{E}(\sigma) = \begin{pmatrix} q_0 & 0 \\ 0 & q_1 \end{pmatrix},
\end{align}
where the output probabilities are derived from the elements of $J(\mathcal{E})$ as $q_0 = r j_{00} + (1-r) j_{22}$ and $q_1 = r j_{11} + (1-r) j_{33}$.

Now apply the vectorization identity $|AXB\rangle\!\rangle = (B^T \otimes A)|X\rangle\!\rangle$ and substituting $A = \sigma^{1/2}$, $X = K_i^\dagger$, and $B = \mathcal{E}(\sigma)^{-1/2}$, we get:
\begin{align}
    |P_m\rangle\!\rangle = \left( \mathcal{E}(\sigma)^{-1/2} \otimes \sigma^{1/2} \right) |K_i^\dagger\rangle\!\rangle \equiv \Delta |K_i^\dagger\rangle\!\rangle,
\end{align}
where we have used the fact that $\mathcal{E}(\sigma)^{-1/2}$ is diagonal and therefore equal to its transpose. Here, $\Delta$ is a diagonal $4 \times 4$ matrix operator:
\begin{align}
    \Delta = \text{diag}\left( \sqrt{\frac{r}{q_0}}, \sqrt{\frac{1-r}{q_0}}, \sqrt{\frac{r}{q_1}}, \sqrt{\frac{1-r}{q_1}} \right).
\end{align}
Since our original Kraus matrices $K_i$ are real, we can express $|K_i^\dagger\rangle\!\rangle = \mathcal{S} |K_i\rangle\!\rangle$, where $\mathcal{S}$ is the SWAP operator over the bipartite space. Using this, the Choi matrix of the Petz map assumes a closed-form matrix structure:
\begin{align}
    J(\mathcal{P}_{\mathcal{E},\sigma}) &= \sum_m \left( \Delta \mathcal{S} |K_i\rangle\!\rangle \right) \left( \langle\!\langle K_i| \mathcal{S} \Delta \right) \nonumber \\
    &= \Delta \mathcal{S} \left( \sum_m |K_i\rangle\!\rangle \langle\!\langle K_i| \right) \mathcal{S} \Delta \nonumber \\
    &= \Delta \mathcal{S} J(\mathcal{E}) \mathcal{S} \Delta.
\end{align}
Applying the swap operator $\mathcal{S}$ to $J(\mathcal{E})$ simply interchanges the inner diagonal elements $j_{11} \leftrightarrow j_{22}$. Multiplying by $\Delta$ on both sides scales the matrix elements. Carrying out this matrix multiplication yields the explicit Choi matrix for the Petz recovery map:
\begin{align}
    J(\mathcal{P}_{\mathcal{E},\sigma}) = 
    \begin{pmatrix}
        \frac{r}{q_0} j_{00} & 0 & 0 & \frac{\sqrt{r(1-r)}}{\sqrt{q_0 q_1}} j_{03} \\
        0 & \frac{1-r}{q_0} j_{22} & \frac{\sqrt{r(1-r)}}{\sqrt{q_0 q_1}} j_{12} & 0 \\
        0 & \frac{\sqrt{r(1-r)}}{\sqrt{q_0 q_1}} j_{12} & \frac{r}{q_1} j_{11} & 0 \\
        \frac{\sqrt{r(1-r)}}{\sqrt{q_0 q_1}} j_{03} & 0 & 0 & \frac{1-r}{q_1} j_{33}
    \end{pmatrix}.
\end{align}

This matrix serves as the target $J(\mathcal{E'})$ when matching against the parameters $p', s', \theta', \kappa', \lambda'$ of the proposed channel construction, where $\mathcal{E'}$ represents the channel in the same family of $\mathcal{E}$ but with these different parameters. 

In the following, we will use $j'_{mn}$ to represent the elements of $J(\mathcal{E'})$ with parameters $p', s', \theta', \kappa', \lambda'$. Let us denote the elements of the target Petz recovery Choi matrix $J(\mathcal{P}_{\mathcal{E},\sigma})$ as $T_{mn}$. From our previous derivation, these target elements are:
\begin{align}
    \begin{split}
        T_{00} &= \frac{r}{q_0} j_{00}, \quad T_{11} = \frac{1-r}{q_0} j_{22}, \\
        T_{22} &= \frac{r}{q_1} j_{11}, \quad T_{33} = \frac{1-r}{q_1} j_{33}, \\
        T_{03} &= \frac{\sqrt{r(1-r)}}{\sqrt{q_0 q_1}} j_{03}, \quad T_{12} = \frac{\sqrt{r(1-r)}}{\sqrt{q_0 q_1}} j_{12}.
    \end{split}
\end{align}

Equating the elements $j'_{mn} = T_{mn}$, we obtain the following equations for the new parameters:
\begin{align}
    j'_{03} &= p' + (1-p')(1-s')\cos^2\frac{\theta'}{2} = T_{03}, \label{eq:match_03} \\
    j'_{12} &= -(1-p')(1-s')\sin^2\frac{\theta'}{2} = T_{12}, \label{eq:match_12} \\
    j'_{00} &= p' + (1-p')(1-s')\cos^2\frac{\theta'}{2} + (1-p')s'\kappa' = T_{00}, \label{eq:match_00} \\
    j'_{22} &= (1-p')(1-s')\sin^2\frac{\theta'}{2} + (1-p')s'\lambda' = T_{22}. \label{eq:match_22}
\end{align}

Since the system is underdetermined as we are fitting a 4-parameter channel family to a specific matrix structure, one parameter remains free. We choose $p'$ as the free parameter and solve for $s', \theta', \kappa', \lambda'$ in terms of $p'$ and the known target elements $T_{mn}$.

First, we solve for the rotation angle $\theta'$. Rearranging \eqref{eq:match_03} and \eqref{eq:match_12}, we have:
\begin{align}
    (1-p')(1-s')\cos^2\frac{\theta'}{2} &= T_{03} - p', \\
    (1-p')(1-s')\sin^2\frac{\theta'}{2} &= -T_{12}.
\end{align}
Dividing the second equation by the first eliminates $s'$, yielding:
\begin{align}
    \tan^2\frac{\theta'}{2} = \frac{-T_{12}}{T_{03} - p'} \quad \implies \quad \theta' = 2 \arctan\left( \sqrt{\frac{-T_{12}}{T_{03} - p'}} \right).
\end{align}

Next, we solve for the scattering probability $s'$. Adding the two rearranged equations gives:
\begin{align}
    (1-p')(1-s')\left(\cos^2\frac{\theta'}{2} + \sin^2\frac{\theta'}{2}\right) &= T_{03} - p' - T_{12} \nonumber \\
    (1-p')(1-s') &= T_{03} - T_{12} - p'.
\end{align}
Isolating $s'$, we find:
\begin{align}
    1 - s' = \frac{T_{03} - T_{12} - p'}{1-p'} \quad \implies \quad s' = \frac{1 - T_{03} + T_{12}}{1-p'}.
\end{align}

Now, we determine the dissipation parameters $\kappa'$ and $\lambda'$. Notice that the first two terms of $j'_{00}$ in \eqref{eq:match_00} are exactly $j'_{03}$. Substituting $j'_{03} = T_{03}$ into \eqref{eq:match_00}:
\begin{align}
    T_{03} + (1-p')s'\kappa' = T_{00} \quad \implies \quad \kappa' = \frac{T_{00} - T_{03}}{(1-p')s'}.
\end{align}
Similarly, the first term of $j'_{22}$ in \eqref{eq:match_22} is exactly $-j'_{12} = -T_{12}$. Substituting this yields:
\begin{align}
    -T_{12} + (1-p')s'\lambda' = T_{22} \quad \implies \quad \lambda' = \frac{T_{22} + T_{12}}{(1-p')s'}.
\end{align}

Finally, substituting the denominator $(1-p')s' = 1 - T_{03} + T_{12}$ into the expressions for $\kappa'$ and $\lambda'$, we get:
\begin{align}
    \label{eq:general solution}
    \begin{aligned}
        \theta' &= 2 \arctan\left( \sqrt{\frac{-T_{12}}{T_{03} - p'}} \right), \\
        s' &= \frac{1 - T_{03} + T_{12}}{1-p'}, \\
        \kappa' &= \frac{T_{00} - T_{03}}{1 - T_{03} + T_{12}}, \\
        \lambda' &= \frac{T_{22} + T_{12}}{1 - T_{03} + T_{12}}.
    \end{aligned}
\end{align}
Interestingly, while $\theta'$ and $s'$ explicitly depend on the choice of the free parameter $p'$, the dissipation parameters $\kappa'$ and $\lambda'$ are independent of $p'$, depending only on the elements of the target Petz recovery matrix.

If such a construction exists, the new parameters need to satisfy the condition $ s',\kappa',\lambda'\in [0,1]$. Working out the validity regions analytically is cumbersome; we do it in the next Appendix for the special case studied in the paper. 

\section{\label{example}The solution of a specific example}
When the parameters of the forward channel are $\theta = \pi/2$, $\kappa = 1$, and $\lambda = 1$, as we have chosen for experiments, the Choi matrix elements $j_{mn}$ are simplified to:
\begin{align}
    j_{00} &= p + \frac{1}{2}(1-p)(1+s), \nonumber \\
    j_{11} &= \frac{1}{2}(1-p)(1-s), \nonumber \\
    j_{22} &= \frac{1}{2}(1-p)(1+s), \nonumber \\
    j_{33} &= p + \frac{1}{2}(1-p)(1-s), \nonumber \\
    j_{03} &= p + \frac{1}{2}(1-p)(1-s), \nonumber \\
    j_{12} &= -\frac{1}{2}(1-p)(1-s).
\end{align}

And the state $\mathcal{E}(\sigma)$ has elements:
\begin{align}
    q_0 &= rp + \frac{1}{2}(1-p)(1+s), \\
    q_1 &= (1-r)p + \frac{1}{2}(1-p)(1-s).
\end{align}
As expected from trace preservation, $q_0 + q_1 = 1$. 

To simplify the subsequent equations, we define a scaling factor $\gamma$:
\begin{align}
    \gamma = \sqrt{\frac{r(1-r)}{q_0 q_1}}.
\end{align}
Then the target elements $T_{mn}$ for the Choi matrix of Petz recovery map take the form:
\begin{align}
    \begin{aligned}
        T_{00} &= \frac{r}{q_0} \left[ p + \frac{1}{2}(1-p)(1+s) \right],  \\
        T_{22} &= \frac{r}{q_1} \left[ \frac{1}{2}(1-p)(1-s) \right],  \\
        T_{03} &= \gamma \left[ p + \frac{1}{2}(1-p)(1-s) \right], \\
        T_{12} &= -\gamma \left[ \frac{1}{2}(1-p)(1-s) \right].
    \end{aligned}
\end{align}

Notice that $T_{03} - T_{12} = \gamma [p + (1-p)(1-s)]$. We can now substitute these expressions into \cref{eq:general solution}. Keeping $p'$ as the free parameter, we obtain the simplified solutions for the specific case $\theta = \pi/2, \kappa = 1, \lambda = 1$:

\begin{align}
    \begin{aligned}
        \theta' &= 2 \arctan\left( \sqrt{ \frac{1}{2}\frac{ \gamma  (1-p)(1-s)  }{ \gamma \left[ p + (1-p)(1-s) \right] - p' } } \right), \\
        s' &= \frac{ 1 - \gamma \left[ p + (1-p)(1-s) \right] }{ 1 - p' }, \\
        \kappa' &= \frac{1}{2}\frac{ (\frac{r}{q_0}-\gamma) \left[2p + (1-p)(1-s) \right] }{ 1 - \gamma \left[ p + (1-p)(1-s) \right] }, \\
        \lambda' &= \frac{1}{2}\frac{ \left( \frac{r}{q_1} - \gamma \right) (1-p)(1-s)  }{ 1 - \gamma \left[ p + (1-p)(1-s) \right] }.
    \end{aligned}
\end{align}

We need to guarantee that there exists at least one choice of our free parameter $p'$ such that the inverse channel parameters represent valid probabilities and angles. Specifically, we require $0 \le s' \le 1$, $0 \le \kappa' \le 1$, $0 \le \lambda' \le 1$, and $\theta' \in \mathbb{R}$. Since $\kappa'$ and $\lambda'$ depend strictly on the elements of the target Petz matrix and are completely independent of $p'$, we can determine the bounds on $r$ starting from $0 \le \lambda' \le 1$. 

We first prove that the denominator of $\lambda'$ is larger than 0:
\begin{align}
    \begin{aligned}
        &1-\gamma [p+(1-p)(1-s)] \geq 0\\
        & \Rightarrow \gamma^2 [p+(1-p)(1-s)]^2 \leq 1\\
        & \Rightarrow r(1-r) [p+(1-p)(1-s)]^2 \leq q_0 q_1 = \frac{1}{4} [1-(-p+2pr-sp+s)^2]\\
        & \Rightarrow 4 r(1-r) [p+(1-p)(1-s)]^2 - [1-(-p+2pr-sp+s)^2] \leq 0
    \end{aligned}
\end{align}
Let $f(s) = 4 r(1-r) [p+(1-p)(1-s)]^2 - [1-(-p+2pr-sp+s)^2] \equiv a s^2 + bs + c$, where $a = (1-p)^2[(2r-1)^2-2] \leq 0$ when $0 \leq p,r \leq 1$. We only need to make sure that $f(s = 0)\geq 0$ and $f(s = 1) \geq 0$:
\begin{align}
    \begin{aligned}
        f(s = 0) &=  (1-p^2) (1-2r)^2 \geq 0;\\
        f(s = 1) &= 4p(1-p)(1-r) \geq 0. 
    \end{aligned}
\end{align}
Therefore, we have proved that the denominator of $\lambda'$ is always no less than 0. Then having $\lambda' \geq 0$ is equal to have
\begin{align}
    \frac{r}{q_1} - \gamma \geq 0. 
\end{align}
Squaring both sides and substituting $\gamma^2 = \frac{r(1-r)}{q_0 q_1}$, we isolate $r$:
\begin{align}
    \begin{aligned}
        & \frac{r^2}{q_1^2} \geq \frac{r(1-r)}{q_0 q_1} \\
        &\Rightarrow r\frac{r}{q_1} -\frac{r(1-r)}{q_0 q_1} \geq 0 \\
        &\Rightarrow \frac{r}{q_0 q_1} \left (r q_0 - (1-r) q_1 \right ) \geq 0\\
        &\Rightarrow \frac{r}{q_0 q_1} (r-q_1) \geq 0 \quad (q_0 = 1-q_1)\\
        &\Rightarrow \frac{r}{q_0 q_1} [r-(1-r)p - \frac{1}{2}(1-p)(1-s)] \geq 0\\
        &\Rightarrow \frac{r}{q_0 q_1} [(1+p)r -p - \frac{1}{2}(1-p)(1-s)] \geq 0 \\
        &\Rightarrow r \geq \frac{1}{2}-\frac{1-p}{2(1+p)}s. 
    \end{aligned}
\end{align}

Next, having $\lambda' \leq 1$ is equal to have:
\begin{align}
    \begin{aligned}
        & T_{22} + T_{12} \leq 1 - T_{03} + T_{12}\\
        &\Rightarrow T_{22} + T_{03} \leq 1\\
        &\Rightarrow T_{03} - T_{33} \leq 0 \quad (T_{22} + T_{33} = 1)\\
        &\Rightarrow \gamma j_{03} - \frac{1-r}{q_1} j_{33} \leq 0\\
        &\Rightarrow (\gamma -\frac{1-r}{q_1}) j_{03} \leq 0 \quad (j_{03} = j_{33} \text{ when } \lambda = 1)\\
        &\Rightarrow \gamma \leq \frac{1-r}{q_1} \quad (j_{03} \geq 0)\\
        &\Rightarrow \frac{r(1-r)}{q_0 q_1} \leq \frac{(1-r)^2}{q_1^2} \\
        &\Rightarrow \frac{r}{q_0} \leq \frac{1-r}{q_1}\\
        &\Rightarrow r q_1 - (1-r)q_0 \leq 0\\
        &\Rightarrow r + q_1 -1 \leq 0 \quad (q_0 = 1-q_1) \\
        &\Rightarrow r + (1-r)p + \frac{1}{2}(1-p)(1-s) - 1 \leq 0\\
        &\Rightarrow r \leq \frac{1+s}{2}. 
    \end{aligned}
\end{align}

To satisfy the remaining conditions—that $s' \in [0, 1]$ and $\theta'$ is a valid real angle—we must be able to choose a free parameter $p'$ such that $p' < T_{03}$ and $p' \le T_{03} - T_{12}$. Since $T_{03} > 0$ and $-T_{12} \ge 0$, it is analytically guaranteed that we can safely select a sufficiently small $p'$ to satisfy these constraints unconditionally for any $r$ inside our derived bounds. 

Finally, we arrive at the strict valid region for the reference state $r$:
\begin{align}
    \frac{1}{2}-\frac{1-p}{2(1+p)}s \leq r \leq \frac{1+s}{2}.
\end{align}
Similarly, by having $0 \leq \kappa' \leq 1$, we get the same region of $r$. 

Having established the valid region for the reference state parameter $r$, we now determine the corresponding allowable range for our free parameter $p'$. For the channel to be physically valid, $p'$ must be a well-defined probability, meaning $0 \le p' \leq 1$. Furthermore, the chosen $p'$ must ensure that the dependent parameters $\theta'$ and $s'$ remain within their physical bounds.

First, for the rotation angle $\theta'$ to be a valid angle, the argument of the square root must be non-negative. From our exact solution, we require:
\begin{align}
    \frac{-T_{12}}{T_{03} - p'} \ge 0.
\end{align}
Since $-T_{12} = \gamma \left[ \frac{1}{2}(1-p)(1-s) \right] \ge 0$, the denominator must be strictly positive to prevent a negative argument or a division by zero. This imposes the strict upper bound:
\begin{align}
    p' < T_{03}.
\end{align}

Next, we verify the condition $0 \le s' \le 1$ for the scattering probability. From our derived solution, $s' = \frac{1 - T_{03} + T_{12}}{1 - p'}$. For $s' \le 1$, we require $1 - T_{03} + T_{12} \le 1 - p'$, which isolates $p'$ as:
\begin{align}
    p' \le T_{03} - T_{12}.
\end{align}
Since $T_{12}$ is strictly non-positive ($-T_{12} \ge 0$), it is  guaranteed that $T_{03} \le T_{03} - T_{12}$. Therefore, the strict bound $p' < T_{03}$ automatically satisfies the condition $s' \le 1$. 

It is also worth noting that the non-negativity of $s'$ (i.e., $s' \geq 0$) requires the numerator to be non-negative, meaning $1- T_{03} + T_{12} \geq 0$. This is intrinsically guaranteed by our previously established valid region for $r$. 

Consequently, the valid region for the free parameter $p'$ is governed entirely by the parameter $T_{03}$, yielding the operational interval $0 \le p' < T_{03}$. Substituting the explicit, fully expanded form of $T_{03}$, we arrive at the final condition for $p'$:
\begin{align}
    0 \leq p' < \frac{\sqrt{(1-r)r}(1+p-s+ps)}{\sqrt{1-(p-2pr+ps-s)^2}}.
\end{align}
As long as the reference state $r$ is chosen within its valid region, any choice of $p'$ in this interval will yield a completely valid set of parameters for the Petz recovery map construction.

\section{\label{sec:steady}Petz recovery map with a steady state as reference}
In this section, we will show that for the forward channel $\mathcal{E}(\rho) = p \rho + (1-p) [ (1-s) \mathcal{R}_{rot}^{\pi/2} (\rho) + s \mathcal{D}_{\kappa = 1,\lambda = 1}(\rho) ]$, choosing its steady state as its reference state, the Petz recovery map is exactly the forward channel itself. When $\kappa = 1, \lambda = 1$, the dissipation part of the forward channel decreases to a fully amplitude damping channel that maps any state to $\ket{0}\bra{0}$, so we can simply the channel as
\begin{align}
    \mathcal{E}(\rho) = p \rho + (1-p) (1-s) \mathcal{R}_{rot}^{\pi/2} (\rho) + (1-p)s \Tr(\rho) \ket{0}\bra{0}. 
\end{align}

First, we will prove that the state $\sigma = r\ket{0}\bra{0} +(1-r)\ket{1}\bra{1}, r = \frac{1+s}{2}$ is the steady state of the forward channel, which is simply prove that $\mathcal{E}(\sigma) = \sigma$. The left hand side equals to 
\begin{align}
    \begin{aligned}
        \text{LHS} &= pr\ket{0}\bra{0} p(1-r) \ket{1}\bra{1} + \frac{(1-p)(1-s)}{2}\ket{0}\bra{0} +  \frac{(1-p)(1-s)}{2}\ket{1}\bra{1} + (1-p)s \ket{0}\bra{0} \\
        &= \frac{1-p+2pr-ps}{2} \ket{0}\bra{0} + \frac{1+p-2pr+ps}{2}\ket{1}\bra{1}\\
        &= \frac{1+s}{2} \ket{0}\bra{0} + \frac{1-s}{2} \ket{1}\bra{1} \quad (r = \frac{1+s}{2})\\
        &= \sigma \\
        &= \text{RHS}. 
    \end{aligned}
\end{align}

Next, we will show that using the steady state as reference state, the Petz recovery map is exactly the forward channel. For steady state, we will have 
\begin{align}
    q_0 = r, q_1 = 1-r. 
\end{align}
Therefore, the coefficient $\gamma = \sqrt{\frac{r(1-r)}{q_0 q_1}} = 1$, and it directly leads to
\begin{align}
    \begin{aligned}
        T_{00} &= \frac{r}{q_0} \left[ p + \frac{1}{2}(1-p)(1+s) \right] = p + \frac{1}{2}(1-p)(1+s) = j_{00},  \\
        T_{22} &= \frac{r}{q_1} \left[ \frac{1}{2}(1-p)(1-s) \right] =  \frac{1}{2}(1-p)(1+s) = j_{22}, \\
        T_{03} &= \gamma \left[ p + \frac{1}{2}(1-p)(1-s) \right] = p + \frac{1}{2}(1-p)(1-s) = j_{03},  \\
        T_{12} &= -\gamma \left[ \frac{1}{2}(1-p)(1-s) \right] = -\frac{1}{2}(1-p)(1-s) = j_{12}.
    \end{aligned}
\end{align}

From above, we have proved that the Petz recovery map with steady state as its reference state is the forward channel.

\section{\label{Exp. simplify}Theoretical derivation for the experimental simplification}

In this part, we will show how we simplify the Petz recovery map to share the same experimental structure as the forward channel. 

\subsection{Rotation operation on the Bloch sphere}
The forward channel is given by:
\begin{align}
    \mathcal{E}(\rho) &= p\rho + (1-p)[\mathcal{R}_{\mathrm{rot}}^{\pi/2}(\rho) + s\mathcal{D}_{\kappa,\lambda}(\rho)]. 
\end{align}

Here, we define $\mathcal{R}_{\mathrm{rot}}^{\theta} = \frac{1}{2}\mathcal{R}_y(\theta)\rho\mathcal{R}_y^{\dagger}(\theta) + \frac{1}{2}\mathcal{R}_y(-\theta)\rho\mathcal{R}_y^{\dagger}(-\theta)$ as the equal mixture of two rotations, $\mathcal{R}_y(\pm\theta)$, about the $y$-axis of the Bloch sphere. A rotation by an angle $\theta$ about an arbitrary axis $\hat{n}$ on the Bloch sphere can be expressed as:
\begin{align}
    \mathcal{R}_{\hat{n}}(\theta) &\equiv \exp(-i\theta \hat{n}\cdot\vec{\sigma}/2) = \cos\frac{\theta}{2}I - i\sin\frac{\theta}{2}(n_x X + n_y Y + n_z Z), \\
    \mathcal{R}_x(\theta) &\equiv \mathrm{e}^{-i\theta X/2} = \cos\frac{\theta}{2}I - i\sin\frac{\theta}{2}X = \begin{pmatrix} \cos\frac{\theta}{2} & -i\sin\frac{\theta}{2} \\ -i\sin\frac{\theta}{2} & \cos\frac{\theta}{2} \end{pmatrix}, \\
    \mathcal{R}_y(\theta) &\equiv \mathrm{e}^{-i\theta Y/2} = \cos\frac{\theta}{2}I - i\sin\frac{\theta}{2}Y = \begin{pmatrix} \cos\frac{\theta}{2} & -\sin\frac{\theta}{2} \\ \sin\frac{\theta}{2} & \cos\frac{\theta}{2} \end{pmatrix}, \\
    \mathcal{R}_z(\theta) &\equiv \mathrm{e}^{-i\theta Z/2} = \cos\frac{\theta}{2}I - i\sin\frac{\theta}{2}Z = \begin{pmatrix} \mathrm{e}^{-i\theta/2} & 0 \\ 0 & \mathrm{e}^{i\theta/2} \end{pmatrix},
\end{align}
where $X$, $Y$, and $Z$ are the standard Pauli matrices. 

Consequently, for $\theta = \pm \pi/2$, we obtain:
\begin{align}
    \mathcal{R}_y\left(\frac{\pi}{2}\right) &= \frac{\sqrt{2}}{2}\begin{pmatrix} 1 & -1 \\ 1 & 1 \end{pmatrix}, \quad \mathcal{R}_y\left(-\frac{\pi}{2}\right) = \frac{\sqrt{2}}{2}\begin{pmatrix} 1 & 1 \\ -1 & 1 \end{pmatrix}.
\end{align}

Let the initial density matrix be $\rho = \begin{pmatrix} a & b \\ c & d \end{pmatrix}$. Applying the equal mixture of $\pm\pi/2$ rotations about the $y$-axis, and utilizing the trace condition $\mathrm{Tr}(\rho) = a + d = 1$, yields:
\begin{align}
    \mathcal{R}_{\mathrm{rot}}^{\frac{\pi}{2}}(\rho) &= \mathcal{R}_y\left(\frac{\pi}{2}\right)\rho\mathcal{R}_y^{\dagger}\left(\frac{\pi}{2}\right) + \mathcal{R}_y\left(-\frac{\pi}{2}\right)\rho\mathcal{R}_y^{\dagger}\left(-\frac{\pi}{2}\right) \nonumber \\
    &= \frac{1}{2}\begin{pmatrix} 1 & -1 \\ 1 & 1 \end{pmatrix}\begin{pmatrix} a & b \\ c & d \end{pmatrix}\begin{pmatrix} 1 & 1 \\ -1 & 1 \end{pmatrix} + \frac{1}{2}\begin{pmatrix} 1 & 1 \\ -1 & 1 \end{pmatrix}\begin{pmatrix} a & b \\ c & d \end{pmatrix}\begin{pmatrix} 1 & -1 \\ 1 & 1 \end{pmatrix} \nonumber \\
    &= \frac{1}{2}\begin{pmatrix} a-c & b-d \\ a+c & b+d \end{pmatrix}\begin{pmatrix} 1 & 1 \\ -1 & 1 \end{pmatrix} + \frac{1}{2}\begin{pmatrix} a+c & b+d \\ c-a & d-b \end{pmatrix}\begin{pmatrix} 1 & -1 \\ 1 & 1 \end{pmatrix} \nonumber \\
    &= \frac{1}{2}\begin{pmatrix} a-c-b+d & a-c+b-d \\ a+c-b-d & a+c+b+d \end{pmatrix} + \frac{1}{2}\begin{pmatrix} a+c+b+d & -a-c+b+d \\ c-a+d-b & a-c+d-b \end{pmatrix} \nonumber \\
    &= \frac{1}{2}\begin{pmatrix} 2a+2d & -2c+2b \\ 2c-2b & 2a+2d \end{pmatrix} \nonumber \\
    &= \begin{pmatrix} 1 & b-c \\ c-b & 1 \end{pmatrix}.
\end{align}

\subsection{Mixture of two arbitrary rotation operations}

Now, we explicitly calculate each component of $\mathcal{R}_{\mathrm{rot}}^\theta(\rho)$. First, evaluating the positive rotation yields:
\begin{align}
\begin{aligned}
    &\mathcal{R}_y(\theta)\rho\mathcal{R}_y^{\dagger}(\theta) = \begin{pmatrix} \cos\frac{\theta}{2} & -\sin\frac{\theta}{2} \\ \sin\frac{\theta}{2} & \cos\frac{\theta}{2} \end{pmatrix}\begin{pmatrix} a & b \\ c & d \end{pmatrix}\begin{pmatrix} \cos\frac{\theta}{2} & \sin\frac{\theta}{2} \\ -\sin\frac{\theta}{2} & \cos\frac{\theta}{2} \end{pmatrix}\\
    &= \begin{pmatrix} a\cos\frac{\theta}{2} - c\sin\frac{\theta}{2} & b\cos\frac{\theta}{2} - d\sin\frac{\theta}{2} \\ a\sin\frac{\theta}{2} + c\cos\frac{\theta}{2} & b\sin\frac{\theta}{2} + d\cos\frac{\theta}{2} \end{pmatrix}\begin{pmatrix} \cos\frac{\theta}{2} & \sin\frac{\theta}{2} \\ -\sin\frac{\theta}{2} & \cos\frac{\theta}{2} \end{pmatrix}\\
    &= \begin{pmatrix} a\cos^2\frac{\theta}{2} - (b+c)\sin\frac{\theta}{2}\cos\frac{\theta}{2} + d\sin^2\frac{\theta}{2} & b\cos^2\frac{\theta}{2} - c\sin^2\frac{\theta}{2} + (a-d)\sin\frac{\theta}{2}\cos\frac{\theta}{2} \\ c\cos^2\frac{\theta}{2} - b\sin^2\frac{\theta}{2} + (a-d)\sin\frac{\theta}{2}\cos\frac{\theta}{2} & a\sin^2\frac{\theta}{2} + (b+c)\sin\frac{\theta}{2}\cos\frac{\theta}{2} + d\cos^2\frac{\theta}{2} \end{pmatrix}\\
    &= \frac{1}{2}\begin{pmatrix} a + a\cos\theta - (b+c)\sin\theta + d - d\cos\theta & (a-d)\sin\theta - c + c\cos\theta + b + b\cos\theta \\ (a-d)\sin\theta + c + c\cos\theta - b + b\cos\theta & a - a\cos\theta + (b+c)\sin\theta + d + d\cos\theta \end{pmatrix}.
\end{aligned}
\end{align}

Similarly, for the negative rotation, we have:
\begin{align}
\begin{aligned}
    &\mathcal{R}_y(-\theta)\rho\mathcal{R}_y^{\dagger}(-\theta) = \begin{pmatrix} \cos\frac{\theta}{2} & \sin\frac{\theta}{2} \\ -\sin\frac{\theta}{2} & \cos\frac{\theta}{2} \end{pmatrix}\begin{pmatrix} a & b \\ c & d \end{pmatrix}\begin{pmatrix} \cos\frac{\theta}{2} & -\sin\frac{\theta}{2} \\ \sin\frac{\theta}{2} & \cos\frac{\theta}{2} \end{pmatrix}\\
    &= \begin{pmatrix} a\cos\frac{\theta}{2} + c\sin\frac{\theta}{2} & b\cos\frac{\theta}{2} + d\sin\frac{\theta}{2} \\ -a\sin\frac{\theta}{2} + c\cos\frac{\theta}{2} & -b\sin\frac{\theta}{2} + d\cos\frac{\theta}{2} \end{pmatrix}\begin{pmatrix} \cos\frac{\theta}{2} & -\sin\frac{\theta}{2} \\ \sin\frac{\theta}{2} & \cos\frac{\theta}{2} \end{pmatrix}\\
    &= \begin{pmatrix} a\cos^2\frac{\theta}{2} + (b+c)\sin\frac{\theta}{2}\cos\frac{\theta}{2} + d\sin^2\frac{\theta}{2} & b\cos^2\frac{\theta}{2} - c\sin^2\frac{\theta}{2} - (a-d)\sin\frac{\theta}{2}\cos\frac{\theta}{2} \\ c\cos^2\frac{\theta}{2} - b\sin^2\frac{\theta}{2} - (a-d)\sin\frac{\theta}{2}\cos\frac{\theta}{2} & a\sin^2\frac{\theta}{2} - (b+c)\sin\frac{\theta}{2}\cos\frac{\theta}{2} + d\cos^2\frac{\theta}{2} \end{pmatrix}\\
    &= \frac{1}{2}\begin{pmatrix} a + a\cos\theta + (b+c)\sin\theta + d - d\cos\theta & -(a-d)\sin\theta - c + c\cos\theta + b + b\cos\theta \\ -(a-d)\sin\theta + c + c\cos\theta - b + b\cos\theta & a - a\cos\theta - (b+c)\sin\theta + d + d\cos\theta \end{pmatrix}.
\end{aligned}
\end{align}

Finally, substituting these expansions into the definition of the mixed rotation channel, $\mathcal{R}_{\mathrm{rot}}^{\theta}(\rho) = \frac{1}{2}\mathcal{R}_y(\theta)\rho\mathcal{R}_y^{\dagger}(\theta) + \frac{1}{2}\mathcal{R}_y(-\theta)\rho\mathcal{R}_y^{\dagger}(-\theta)$, we obtain:
\begin{align}
\begin{aligned}
    \mathcal{R}_{\mathrm{rot}}^{\theta}(\rho) &= \frac{1}{2} \left[ \mathcal{R}_y(\theta)\,\rho\,\mathcal{R}_y^{\dagger}(\theta) + \mathcal{R}_y(-\theta)\,\rho\,\mathcal{R}_y^{\dagger}(-\theta) \right] \\
    &= \frac{1}{2} \begin{pmatrix} a + a\cos\theta + d - d\cos\theta & b + b\cos\theta - c + c\cos\theta \\ -b + b\cos\theta + c + c\cos\theta & a - a\cos\theta + d + d\cos\theta \end{pmatrix}\\
    &= \begin{pmatrix} \frac{a+a\cos\theta}{2} + \frac{d-d\cos\theta}{2} & \frac{b+b\cos\theta-(c-c\cos\theta)}{2} \\ \frac{-(b-b\cos\theta)+c+c\cos\theta}{2} & \frac{a-a\cos\theta}{2} + \frac{d+d\cos\theta}{2} \end{pmatrix}\\
    &= \begin{pmatrix} a\cos^2\frac{\theta}{2} + d\sin^2\frac{\theta}{2} & b\cos^2\frac{\theta}{2} - c\sin^2\frac{\theta}{2} \\ -b\sin^2\frac{\theta}{2} + c\cos^2\frac{\theta}{2} & a\sin^2\frac{\theta}{2} + d\cos^2\frac{\theta}{2} \end{pmatrix}. \label{eq:general_rotation}
\end{aligned}
\end{align}

\subsection{Experimental implementation for the rotation channel}

The Jones matrix for a quarter-wave plate (QWP) oriented at an angle $\theta$ is given by:
\begin{align}
    J_{\mathrm{QWP}}(\theta) &= \begin{pmatrix} \cos^2\theta + i\sin^2\theta & (1-i)\sin\theta \cos\theta \\ (1-i)\sin\theta \cos\theta & \sin^2\theta + i\cos^2\theta \end{pmatrix}.
\end{align}

Specifically, for orientation angles of $\theta = \pm 45^{\circ}$, the matrices simplify to:
\begin{align}
    J_{\mathrm{QWP}}(45^{\circ}) &= \begin{pmatrix} \frac{1+i}{2} & \frac{1-i}{2} \\ \frac{1-i}{2} & \frac{1+i}{2} \end{pmatrix} = \frac{1+i}{2}\begin{pmatrix} 1 & -i \\ -i & 1 \end{pmatrix}, \\
    J_{\mathrm{QWP}}(-45^{\circ}) &= \begin{pmatrix} \frac{1+i}{2} & \frac{i-1}{2} \\ \frac{i-1}{2} & \frac{1+i}{2} \end{pmatrix} = \frac{1+i}{2}\begin{pmatrix} 1 & i \\ i & 1 \end{pmatrix}.
\end{align}

Meanwhile, the operator for a rotation about the $x$-axis of the Bloch sphere by an angle $\theta$ is defined as:
\begin{align}
    \mathcal{R}_x(\theta) &= \begin{pmatrix} \cos\frac{\theta}{2} & -i\sin\frac{\theta}{2} \\ -i\sin\frac{\theta}{2} & \cos\frac{\theta}{2} \end{pmatrix}.
\end{align}

For rotation angles of $\theta = \pm \pi/2$, this evaluates to:
\begin{align}
    \mathcal{R}_x\left(\frac{\pi}{2}\right) &= \begin{pmatrix} \frac{\sqrt{2}}{2} & -i\frac{\sqrt{2}}{2} \\ -i\frac{\sqrt{2}}{2} & \frac{\sqrt{2}}{2} \end{pmatrix} = \frac{\sqrt{2}}{2}\begin{pmatrix} 1 & -i \\ -i & 1 \end{pmatrix}, \\
    \mathcal{R}_x\left(-\frac{\pi}{2}\right) &= \begin{pmatrix} \frac{\sqrt{2}}{2} & i\frac{\sqrt{2}}{2} \\ i\frac{\sqrt{2}}{2} & \frac{\sqrt{2}}{2} \end{pmatrix} = \frac{\sqrt{2}}{2}\begin{pmatrix} 1 & i \\ i & 1 \end{pmatrix}.
\end{align}

Comparing the operators, we see that $J_{\mathrm{QWP}}(\pm 45^{\circ})$ and $\mathcal{R}_x(\pm \pi/2)$ differ only by a global phase of $e^{i\pi/4} = (1+i)/\sqrt{2}$. Since density matrices are invariant under global phases, a $\pm \pi/2$ rotation on the Bloch sphere about the $x$-axis can be physically implemented using a quarter-wave plate oriented at $\pm 45^{\circ}$. 

Let the initial density matrix be $\rho = \begin{pmatrix} a & b \\ c & d \end{pmatrix}$. Applying the $\pi/2$ rotation about the $x$-axis (via the $45^{\circ}$ QWP), the state transforms as:
\begin{align}
    \mathcal{R}_x\left(\frac{\pi}{2}\right)\rho\mathcal{R}_x^{\dagger}\left(\frac{\pi}{2}\right) &= \frac{1}{2}\begin{pmatrix} 1 & -i \\ -i & 1 \end{pmatrix}\begin{pmatrix} a & b \\ c & d \end{pmatrix}\begin{pmatrix} 1 & i \\ i & 1 \end{pmatrix} = \frac{1}{2}\begin{pmatrix} a+d+(b-c)i & b+c+(a-d)i \\ b+c+(d-a)i & a+d+(c-b)i \end{pmatrix}.
\end{align}

Next, passing this intermediate state through a phase damping channel characterized by a decoherence parameter $L$ yields:
\begin{align}
    \rho_{\mathrm{damp}} &= \frac{1}{2}\begin{pmatrix} a+d+(b-c)i & (b+c+(a-d)i)e^{-L} \\ (b+c+(d-a)i)e^{-L} & a+d+(c-b)i \end{pmatrix}.
\end{align}

Subsequently, applying a $-\pi/2$ rotation about the $x$-axis (via the $-45^{\circ}$ QWP) to $\rho_{\mathrm{damp}}$ produces the final output state:
\begin{align}
    \mathcal{R}_x\left(-\frac{\pi}{2}\right)\rho_{\mathrm{damp}}\mathcal{R}_x^{\dagger}\left(-\frac{\pi}{2}\right) &= \frac{1}{2}\begin{pmatrix} a+d+(a-d)e^{-L} & b-c+(b+c)e^{-L} \\ c-b+(b+c)e^{-L} & a+d+(d-a)e^{-L} \end{pmatrix}. \label{eq:qwp_final_state}
\end{align}

Comparing \cref{eq:qwp_final_state} with \cref{eq:general_rotation}, we observe that the decoherence factor maps directly to the rotation angle such that $e^{-L} = \cos\theta$. Therefore, we can simulate an arbitrary rotation angle within this framework simply by tuning the decoherence parameter $L$ of the phase plate.

\subsection{Experimental simplification for the forward channel}

Recall that the original form of the forward channel is expressed as:
\begin{equation}
    \mathcal{E}(\rho) = p\rho + (1-p)[(1-s)\mathcal{R}_{\mathrm{rot}}^{\pi/2}(\rho) + s\mathcal{D}_{\kappa,\lambda}(\rho)].
\end{equation}

From our previous derivation, the equal mixture of $\pm\pi/2$ rotations evaluates to:
\begin{equation}
    \mathcal{R}_{\mathrm{rot}}^{\frac{\pi}{2}}(\rho) = \frac{1}{2}\begin{pmatrix} a+d & b-c \\ c-b & a+d \end{pmatrix}.
\end{equation}

The unitary portion of the channel, $p\rho + (1-p)(1-s)\mathcal{R}_{\mathrm{rot}}^{\pi /2}(\rho)$, can be recast as an effective rotation channel $\mathcal{R}_{\mathrm{rot}}^{\alpha}(\rho)$ with a specific angle $\alpha$. Expanding this term yields:
\begin{equation} \label{eq:simplified_mixture}
\begin{aligned}
    &p\rho + (1-p)(1-s)\mathcal{R}_{\mathrm{rot}}^{\frac{\pi}{2}}(\rho) \\
    &= p\begin{pmatrix} a & b \\ c & d \end{pmatrix} + \frac{(1-s)(1-p)}{2}\begin{pmatrix} a+d & b-c \\ c-b & a+d \end{pmatrix} \\
    &= \begin{pmatrix} \left[\frac{(1-s)(1-p)}{2}+p\right]a + \frac{(1-s)(1-p)}{2}d & \left[\frac{(1-s)(1-p)}{2}+p\right]b - \frac{(1-s)(1-p)}{2}c \\ \left[\frac{(1-s)(1-p)}{2}+p\right]c - \frac{(1-s)(1-p)}{2}b & \left[\frac{(1-s)(1-p)}{2}+p\right]d + \frac{(1-s)(1-p)}{2}a \end{pmatrix}.
\end{aligned}
\end{equation}

Comparing this result with the general rotation matrix in \cref{eq:general_rotation}, we deduce the mapping relations for the effective rotation angle $\alpha$:
\begin{equation}
\begin{cases}
    \dfrac{(1-s)(1-p)}{2} + p = \cos^2\dfrac{\alpha}{2}, \\[10pt]
    \dfrac{(1-s)(1-p)}{2} = \sin^2\dfrac{\alpha}{2}.
\end{cases}
\end{equation}

Notice that the sum of the coefficients on the left-hand side is $p + (1-p)(1-s) = 1 - s + ps$. Since this sum is not equal to $1$, we must extract this normalization factor. The renormalized expression becomes:
\begin{equation}
\begin{aligned}
    p\rho + (1-p)(1-s)\mathcal{R}_{\mathrm{rot}}^{\pi/2}(\rho) &= (1-s+ps)\mathcal{R}_{\mathrm{rot}}^{\alpha}(\rho) \\
    &= (1-s+ps)\begin{pmatrix} a\cos^2\frac{\alpha}{2} + d\sin^2\frac{\alpha}{2} & b\cos^2\frac{\alpha}{2} - c\sin^2\frac{\alpha}{2} \\ -b\sin^2\frac{\alpha}{2} + c\cos^2\frac{\alpha}{2} & a\sin^2\frac{\alpha}{2} + d\cos^2\frac{\alpha}{2} \end{pmatrix}.
\end{aligned}
\end{equation}

Substituting this back into the expression for $\mathcal{E}(\rho)$, the quantum channel simplifies to:
\begin{equation}
\begin{aligned}
    \mathcal{E}(\rho) &= p\rho + (1-p)\left[ (1-s)\mathcal{R}_{\mathrm{rot}}^{\pi/2}(\rho) + s\mathcal{D}_{\kappa,\lambda}(\rho) \right] \\
    &= (1-s+ps)\mathcal{R}_{\mathrm{rot}}^{\alpha}(\rho) + (1-p)s\mathcal{D}_{\kappa,\lambda}(\rho) \\
    &= (1-x)\mathcal{R}_{\mathrm{rot}}^{\alpha}(\rho) + x\mathcal{D}_{\kappa,\lambda}(\rho),
\end{aligned}
\end{equation}
where we have introduced the effective damping parameter $x = (1-p)s$.

We now extend this simplification approach to the Petz recovery map, denoted as $\mathcal{P}_{\mathcal{E},\sigma}(\rho)$, which takes a similar form but is parameterized by $p'$, $s'$, an arbitrary rotation angle $\theta'$, and dissipation parameters $\kappa', \lambda'$:
\begin{equation}
    \mathcal{P}_{\mathcal{E},\sigma}(\rho) = p'\rho + (1-p')\left[ (1-s')\mathcal{R}_{\mathrm{rot}}^{\theta'}(\rho) + s'\mathcal{D}_{\kappa',\lambda'}(\rho) \right].
\end{equation}
Recall that the generalized rotation channel is defined as $\mathcal{R}_{\mathrm{rot}}^{\theta}(\rho) = \frac{1}{2}\left[\mathcal{R}_y(\theta)\rho\mathcal{R}_y^\dagger(\theta) + \mathcal{R}_y(-\theta)\rho\mathcal{R}_y^\dagger(-\theta)\right]$. Focusing on the unitary mixture portion of $\mathcal{P}_{\mathcal{E},\sigma}(\rho)$, we expand it as follows:
\begin{equation}
\begin{aligned}
    &p'\rho + (1-p')(1-s')\mathcal{R}_{\mathrm{rot}}^{\theta'}(\rho) \\
    &= \begin{pmatrix} p'a & p'b \\ p'c & p'd \end{pmatrix} + (1-p')(1-s')\begin{pmatrix} a\cos^2\frac{\theta'}{2} + d\sin^2\frac{\theta'}{2} & b\cos^2\frac{\theta'}{2} - c\sin^2\frac{\theta'}{2} \\ c\cos^2\frac{\theta'}{2} - b\sin^2\frac{\theta'}{2} & a\sin^2\frac{\theta'}{2} + d\cos^2\frac{\theta'}{2} \end{pmatrix} \\
    &= (1-s'+s'p')\begin{pmatrix} a\cos^2\frac{\alpha'}{2} + d\sin^2\frac{\alpha'}{2} & b\cos^2\frac{\alpha'}{2} - c\sin^2\frac{\alpha'}{2} \\ -b\sin^2\frac{\alpha'}{2} + c\cos^2\frac{\alpha'}{2} & a\sin^2\frac{\alpha'}{2} + d\cos^2\frac{\alpha'}{2} \end{pmatrix}.
\end{aligned}
\end{equation}

The total probability weight for this segment is $p' + (1-p')(1-s')\left(\cos^2\frac{\theta'}{2} + \sin^2\frac{\theta'}{2}\right) = 1 - s' + s'p'$. Since this weight is not equal to $1$, we normalize the terms to establish the mapping relation to a new effective angle $\alpha'$:
\begin{equation}
\begin{cases}
    \dfrac{1}{1-s'+s'p'}\left[ p' + \dfrac{(1-s')(1-p')}{2}\cos^2\frac{\theta'}{2} \right] = \cos^2\dfrac{\alpha'}{2}, \\[12pt]
    \dfrac{1}{1-s'+s'p'}\left[ \dfrac{(1-s')(1-p')}{2}\sin^2\frac{\theta'}{2} \right] = \sin^2\dfrac{\alpha'}{2}.
\end{cases}
\end{equation}

Using this effective angle, the recovery map $\mathcal{P}_{\mathcal{E},\sigma}(\rho)$ shares the exact same structure as the simplified forward channel:
\begin{equation}
\begin{aligned}
    \mathcal{P}_{\mathcal{E},\sigma}(\rho) &= p'\rho + (1-p')\left[ (1-s')\mathcal{R}_{\mathrm{rot}}^{\theta'}(\rho) + s'\mathcal{D}_{\kappa',\lambda'}(\rho) \right] \\
    &= (1-s'+s'p')\mathcal{R}_{\mathrm{rot}}^{\alpha'}(\rho) + (1-p')s'\mathcal{D}_{\kappa',\lambda'}(\rho) \\
    &= (1-x')\mathcal{R}_{\mathrm{rot}}^{\alpha'}(\rho) + x'\mathcal{D}_{\kappa',\lambda'}(\rho),
\end{aligned}
\end{equation}
where the effective parameters $x'$ and $\alpha'$ satisfy the following system of equations:
\begin{equation}
\begin{cases}
    p' + \dfrac{(1-s')(1-p')}{2}\cos\theta' = \cos\alpha' \, (1-x'), \\
    x' = (1-p')s'.
\end{cases}
\end{equation}

Inverting this system allows us to solve for the target channel parameters $p'$ and $s'$ in terms of the effective experimental parameters $x$, $\alpha$, and the underlying angle $\theta$:
\begin{equation}
\begin{cases}
    p' = \dfrac{2(1-x')\cos\alpha' + x'\cos\theta' - \cos\theta'}{2 - \cos\theta'}, \\
    s' = \dfrac{2x' - x'\cos\theta'}{2 + x'\cos\theta' - 2\cos\theta'}.
\end{cases}
\end{equation}

For the experimental implementation, this effective rotation angle $\alpha'$ is physically governed by the decoherence factor of the phase plate, satisfying the relation $e^{-L'} = \cos\alpha'$.

\section{\label{Exp. detail}Experimental implementation of Petz recovery map}

In this part, we will go through the details of the experimental implementation of the forward channel and its Petz recovery map. The experimental setup will introduce decoherence between two orthogonal polarizations to achieve the corresponding channels. Here we use Liquid Crystal Phase Retarder (LCPR) to introduce the time delay between two orthogonal polarizations. The LCPR can be controlled by the applied voltage, which can change the refractive index of the liquid crystal material inside the LCPR.

In the experiment, the angle $\alpha$ of the rotation channel $\mathcal{R}_{\text{rot}}^{\alpha}$ can be controlled by tuning the decoherence length by using phase retarder with different thicknesses. The rotation channel $\mathcal{R}_{\text{rot}}^{\alpha}$ can be achieved by the combination of three operations $\mathcal{R}_{x}(\alpha)\mathcal{D}_{L}\mathcal{R}_{x}(-\alpha)$ in the experiment, where $\mathcal{R}_x(\pm \alpha)$ is the rotation of angle $\alpha$ around the $x-$axis of the Bloch sphere and $\mathcal{D}_{L}$ is the dephasing operations where $L$ is the coherence length.  In this way, we can control the degree of the decoherence $e^{-L}$ to control the rotation value $\cos \alpha $.

Consider two pure qubit states that share the same population amplitudes $a, b \geq 0$ (with $a^2 + b^2 = 1$) but differ in relative phase:
\begin{equation}\label{eq:states}
  \ket{\psi_1} = \begin{pmatrix} a \\ b\,e^{i\phi} \end{pmatrix}, \qquad
  \ket{\psi_2} = \begin{pmatrix} a \\ b\,e^{i\psi} \end{pmatrix}.
\end{equation}
Their corresponding density matrices are
\begin{equation}
  \ket{\psi_k}\!\bra{\psi_k} = \begin{pmatrix} a^2 & ab\,e^{-i\theta_k} \\ ab\,e^{i\theta_k} & b^2 \end{pmatrix},
\end{equation}
where $\theta_1 = \phi$ and $\theta_2 = \psi$.

An incoherent mixture with weights $\eta$ and $1-\eta$ (where $0 \leq \eta \leq 1$) yields the density matrix
\begin{equation}\label{eq:mixture}
  \rho = (1-\eta)\,\ket{\psi_1}\!\bra{\psi_1} + \eta\,\ket{\psi_2}\!\bra{\psi_2}
  = \begin{pmatrix}
    a^2 & ab\,\alpha^{*} \\[2pt]
    ab\,\alpha & b^2
  \end{pmatrix},
\end{equation}
where we define
\begin{equation}\label{eq:alpha}
  \alpha \equiv (1-\eta)\,e^{i\phi} + \eta\,e^{i\psi}.
\end{equation}
Separating into real and imaginary parts gives
\begin{align}
  \operatorname{Re}[\rho_{01}] &= ab\bigl[(1-\eta)\cos\phi + \eta\cos\psi\bigr], \label{eq:re} \\
  \operatorname{Im}[\rho_{01}] &= -ab\bigl[(1-\eta)\sin\phi + \eta\sin\psi\bigr]. \label{eq:im}
\end{align}
\cref{eq:re}--\cref{eq:im} establish the mapping between the mixing weight $\eta$ and the off-diagonal element $\rho_{01}$ of an arbitrary qubit state with diagonal elements $a^2$ and $b^2$.

To prepare a diagonal state with $\rho_{01} = 0$, we require $\alpha = 0$ in Eq.~\eqref{eq:alpha}. Taking the modulus gives $(1-\eta) = \eta$, i.e.\ $\eta = 1/2$. Substituting back:
\begin{equation}
  \tfrac{1}{2}\,e^{i\phi} + \tfrac{1}{2}\,e^{i\psi} = 0
  \quad \Longrightarrow \quad
  e^{i(\phi - \psi)} = -1
  \quad \Longrightarrow \quad
  \phi - \psi = \pi \pmod{2\pi}.
\end{equation}
Therefore, a maximally dephased (diagonal) state $\rho = \operatorname{diag}(a^2,\, b^2)$ is obtained by equally mixing two pure states whose relative phases differ by $\pi$.

\end{document}